\documentclass[prb,superscriptaddress]{revtex4}

\usepackage{epsf}
\usepackage{amsmath}

\def\Tr{\mathop{\rm Tr}\nolimits}
\def\const{{\rm const}}
\def\be{\begin{equation}}
\def\ee{\end{equation}}
\newcommand{\corr}[1]{\langle #1\rangle}
\def\vp{\varphi}

\begin{document}

\title{Quantum mechanics with a time-dependent random unitary Hamiltonian:\\
A perturbative study of the nonlinear Keldysh sigma-model}

\author{D. A. Ivanov}
\affiliation{Ecole Polytechnique Federale de Lausanne (EPFL),
Institute for Theoretical Physics,
CH-1015 Lausanne, Switzerland}

\author{M. A. Skvortsov}
\affiliation{L. D. Landau Institute for Theoretical Physics, Moscow 119334, Russia}

\date{March 29, 2006}

\begin{abstract}
We analyze the perturbative series of the Keldysh-type sigma-model
proposed recently
for describing the quantum mechanics with
time-dependent Hamiltonians from the unitary Wigner--Dyson random-matrix
ensemble. We observe that vertices of orders higher than four cancel, which
allows us to reduce the calculation of the energy-diffusion constant to
that in a special kind of the matrix $\phi^4$ model. We further verify that the perturbative
four-loop correction to the energy-diffusion constant in the high-velocity
limit cancels, in agreement with the conjecture of one of the authors.
\end{abstract}

\maketitle

\section{Introduction}

While spectral properties of static random-matrix Hamiltonians have been
studied in detail by various methods \cite{Mehta,Guhr,Efetov-book},
much less is known about
quantum-mechanical evolution with a {\it time-dependent} random-matrix
Hamiltonian:
\be
  i\frac{\partial}{\partial t} \Psi = H(t) \Psi.
\label{schroedinger}
\ee
Such a system with the Hamiltonian $H(t)$ belonging to one of the
three Wigner--Dyson random-matrix ensembles (unitary, orthogonal or
symplectic) has been studied by Wilkinson in Ref.~\onlinecite{wilkinson}.
The trajectory of the Hamiltonian $H(t)$ in the space of Hermitian matrices
is assumed to be nearly linear on the time scales relevant for the problem.
This assumption is justified in the limit of large matrix
dimension $N$: in this limit the energy level spacing $\Delta$ is small,
and a small variation of the Hamiltonian matrix elements (of order $\Delta$,
which is much smaller than the matrix elements themselves)
already shifts the energy levels by the order of the level spacing
and thus changes the level correlations completely. Therefore the
relevant time scales are small in $N$, and any trajectory $H(t)$ with
smooth (independent of $N$) time dependence may be
approximated as linear. This is the usual reasoning in the studies of
parametric level statistics \cite{SimonsAltshuler93} which deduces that
under such an assumption the spectral correlations acquire universal
properties (independent of the particular choice of the trajectory).

Following the traditional notation and for the future possibility of
describing different time dependencies of $H(t)$,
we introduce the time dependence of the Hamiltonian in two steps. First
we define the class of linear trajectories $H(\varphi)$ and
then let the parameter $\varphi$ be a given function of time $t$.
In the present paper we restrict our discussion
to the Gaussian Unitary Ensemble, for which the linear trajectories
$H(\varphi)$ may be defined by the pair-correlation functions
\begin{subequations}
\begin{gather}
  \overline{H_{ij}(\vp) H_{kl}(\vp)}
  = \frac{N\Delta^2}{\pi^2} \, \delta_{il} \delta_{jk} ,
\label{GUE}
\\
  \overline{\big(H(\vp)-H(\vp')\big)_{ij} \big(H(\vp)-H(\vp')\big)_{kl}}
  =  \delta_{il} \delta_{jk} (\vp-\vp')^2 \Delta^2 C(0)
  + O\left( \frac{(\vp-\vp')^4}{N}\right) ,
\label{HH}
\end{gather}
\end{subequations}
where $\Delta$ is the mean level spacing in the center
of the Wigner semicircle, and $C(0)$ is the conventional
notation for the sensitivity of the energy spectrum on the parameter
$\vp$ (see, e.g., Ref.~\onlinecite{SimonsAltshuler93}).

Two possibilities of the motion along the trajectory $H(\vp)$
are of principal importance:
\begin{itemize}
\item[(i)]
Linear time dependence $\vp(t)=vt$. This is the problem studied
by Wilkinson \cite{wilkinson} and also the situation considered in the present
paper (except in the Section \ref{section-phi(t)} where more general
time dependencies $\vp(t)$ are discussed).
\item[(ii)]
Periodic time dependence $\vp(t)=\cos(\omega t)$. In that case,
diffusion in the energy space is suppressed by quantum interference,
leading to the phenomenon of
{\em dynamic localization}~\cite{Casati79,wilkinson2,BSK03}.
We do not discuss the periodic problem in the present paper.
\end{itemize}

We further specify to the case of linear motion along the trajectory
$\vp(t)=vt$ and replace the parameter $C(0)$ by a more convenient for
our present discussion dimensionless parameter $\alpha$ defined as
\begin{equation}
\alpha
= \frac{\pi}{\Delta^2} C(0) v^2
= \frac{\pi}{\Delta^4}
\overline{\left( \frac{\partial E_n}{\partial t}\right)^2}
.
\label{alpha-definition}
\end{equation}
Depending on whether $\alpha$ is much smaller or
much larger than one, the transitions between levels may be described either
as Landau--Zener transitions between the neighboring levels
or as transitions in the continuum spectrum
according to the linear-response Kubo formula. In both limits, the
quantum-mechanical state effectively experiences a diffusion in energy,
so that the energy drift over a large time $T$ is given by
\begin{equation}
\overline{ [E(T)-E(0)]^2 } = 2 D\, T\, \Delta^3\, .
\label{diffusion-definition}
\end{equation}
The dimensionless diffusion coefficient $D$ depends on $\alpha$. A remarkable result of
Wilkinson~\cite{wilkinson} is that in the case of unitary random-matrix ensemble,
in both limits of {\em large}\/ and {\em small}\/ $\alpha$,
the energy-diffusion
coefficient is given by the same expression
\begin{equation}
D(\alpha)=\alpha.
\end{equation}

Recently, this problem has been analyzed further
by one of the authors \cite{skvor,SBK04} with
the use of a $\sigma$-model constructed from the Keldysh-integral averaging
over random matrices $H(t)$. The $\sigma$-model formulation
of Refs.~\onlinecite{skvor}, \onlinecite{SBK04}
contains, in principle, full information about the function $D(\alpha)$, and
its diagrammatic expansion allows us to compute the perturbative series
for $D(\alpha)$ in the limit of large $\alpha$.
For the driven Gaussian {\em orthogonal}\/ matrices, the one-loop correction
has the form~\cite{skvor,GOE-comment}:
$D(\alpha)=\alpha(1+d_1^{\rm(O)}\pi^{-1}\alpha^{-1/3}+\dots)$,
where $d_1^{\rm(O)}=\Gamma(1/3)\,3^{-2/3}$.
For the {\em unitary}\/ ensemble,
the number of loops in the diagrams for calculating $D(\alpha)$ must be
even, which corresponds to expanding in powers of $\alpha^{-2/3}$:
\be
  D(\alpha) = \alpha
  \left(
    1 + \frac{d_2}{\pi^2\alpha^{2/3}} + \frac{d_4}{\pi^4\alpha^{4/3}} + \dots
  \right) ,
\label{D-series}
\ee
where we took into account that, according to Sec.~\ref{section-diagrams-D},
the expansion parameter is $\pi^{-2}\alpha^{-2/3}$.
In Refs. \onlinecite{skvor} and \onlinecite{SBK04},
it has been found that $d_2=0$, and it has been further
conjectured that all higher-order perturbative terms also vanish.
In the present paper we shall verify
that this conjecture holds up to the four-loop order: the
value of $d_4$ obtained by numerical evaluation
vanishes with very high accuracy ($|d_4|<3\times10^{-4}$)
which is a strong argument in favor of
\be
  d_4 = 0.
\ee

In the process of developing the diagrammatic expansion for
$D(\alpha)$ we observe a remarkable cancellation of diagrammatic
vertices of order higher than four. This cancellation is proven to
{\it all} orders in the perturbation theory by combinatoric means.
Therefore we find that, at the level of perturbative series, the
non-linear $\sigma$-model of Refs.~\onlinecite{skvor,SBK04}
is exactly equivalent to a matrix $\phi^4$-type theory.
We show that the resulting theory can be obtained from the
initial $\sigma$-model by applying a transformation analogous
to the Dyson-Maleev~\cite{Dyson56,Maleev57} parameterization of quantum spin operators.
The equivalence of the $\sigma$-model to a kind of a $\phi^4$-type theory
may be a more important result than just a tool
for calculating higher-order corrections in (\ref{D-series}): in
particular, this equivalence between two theories
may have its counterparts for other types of non-linear $\sigma$-models,
such as the one describing the diffusion of a quantum particle in
a disordered media~\cite{Wegner1979,ELK1980,Efetov-book}.
We perform one straightforward generalization
of this diagram cancellation to the case of
arbitrary time-dependence of the control parameter $\vp(t)$.

The rest of the paper is organized as follows. In Section \ref{section-model}
we describe the $\sigma$-model of Refs.~\onlinecite{skvor,SBK04}. In Section
\ref{section-diagrams}, we develop the rules of the diagrammatic expansion.
Further in Section \ref{section-diffuson} we apply these rules for
constructing the perturbative series for $D(\alpha)$.
In Section \ref{section-cancellation} we prove the cancellation of
the vertices of order higher than four in the ``rational''
parameterization. We further formulate the new diagrammatic rules
and the corresponding $\phi^4$ theory. In the following Section
\ref{section-diagrams-D}, we apply the derived equivalence to
explicitly write down the corrections to the diffusion coefficient
(\ref{D-series}). The numerical evaluation of the diagrams to the
four-loop order is reported in Section \ref{section-four-loops}.
In Section \ref{section-phi(t)} we generalize our results to a more
general situation of arbitrary dependence of $\vp(t)$.
In Section \ref{section-DM} we demonstrate that our $\phi^4$ theory
can be obtained from the initial $\sigma$-model by transformation
similar to the Dyson-Maleev transformation.
We
conclude our discussion in Section \ref{section-conclusion}.
Technically complicated details of the calculations are delegated
to Appendices.

\section{Keldysh sigma-model}
\label{section-model}

The starting point of our analysis is the $\sigma$-model action
derived in Ref.~\onlinecite{skvor} for the unitary ensemble. The
field variable is the operator $Q$ which is the integral kernel in
time domain with values in $2\times 2$ matrices in
retarded-advanced Keldysh space. Technically, we write $Q$ as a
$2\times 2$ matrix $Q_{tt'}$ depending on time variables $t$ and $t'$.
The general expression for the action with arbitrary $\vp(t)$
has the form
\be
  S[Q] =
  \frac{\pi i}{\Delta} \Tr \hat{E}Q
  - \frac{\pi^2C(0)}{4} \Tr [\hat\vp,Q]^2 .
\label{action-general}
\ee
Here ${\hat E}$ is the energy operator with the matrix
elements ${\hat E}_{tt'} = i \delta_{tt'} \partial_{t'}$,
the diagonal in time representation operator $\hat\vp$
is defined by the matrix elements $\hat\vp_{tt'} = \delta_{tt'} \vp(t')$,
and the trace is taken both over the Keldysh and time spaces.
The commutator in the last term of Eq.~(\ref{action-general})
vanishes if $\vp(t)=\const$, and is generally nonzero for
a time-dependent $\vp(t)$.

To study the system's dynamics for the case of the linear
perturbation $\vp(t)=vt$, we prefer to simplify the notation
by measuring time in the units of $\Delta^{-1}$.
The resulting action in
dimensionless units may be written as
\begin{subequations}
\label{action-sigma-model}
\be
  S[Q]= S_{\rm E}[Q] + S_{\rm kin}[Q],
\label{action-total}
\ee
where
\begin{gather}
  S_{\rm E}[Q] = \pi i \Tr_{{\rm K},t} (\hat{E}Q) = -\frac{\pi}{2} \Tr_{\rm K}
\int dt\, (\partial_1 - \partial_2) Q_{t_1 t_2} \Big|_{t_1=t_2=t} ,
\label{action-E}
\\
  S_{\rm kin}[Q]
  = -\frac{\pi\alpha}{4} \Tr_{{\rm K},t} [t,Q]^2
  = \frac{\pi\alpha}{4}
    \Tr_{\rm K} \int\!\!\int dt\, dt'\, (t-t')^2\, Q_{tt'} Q_{t't} .
\label{action-kin}
\end{gather}
\end{subequations}
Here $\Tr_{{\rm K},t}$ and $\Tr_{\rm K}$ denote the traces over the
full Keldysh-time space and over the two-dimensional Keldysh space only,
and $\alpha$ is the same
dimensionless coupling constant as in (\ref{alpha-definition}) and
(\ref{D-series}).

The $Q$-matrix itself is subject to the constraint
${Q}^2=1$, or, more precisely,
\begin{equation}
  \int dt'\, Q_{t t'} Q_{t' t''}
  =
  \begin{pmatrix}
    \delta^R_{t t''} & 0 \\
    0 & \delta^A_{t t''}
  \end{pmatrix} .
\label{constraint}
\end{equation}
Where we have introduced the ``retarded'' and ``advanced''
$\delta$-functions $\delta^{R,A}_{t_1,t_2}=\delta(t_1-t_2 \mp \varepsilon)$
with an infinitesimal shift $\varepsilon$. This
definition ensures the proper regularization~\cite{KamenevAndreev99}
of the functional integral of $\exp(-S[Q])$.

The functional integration in $Q$ is performed over an appropriate
real submanifold of the complex manifold defined by the constraint
(\ref{constraint}) (this procedure is standard
in the sigma-model derivation
both in Keldysh~\cite{KamenevAndreev99,HorbachSchoen1993}
and supersymmetric~\cite{Efetov-book} formalisms).
We take this manifold to be the orbit
\begin{equation}
Q=U^{-1} \Lambda U
\label{orbit}
\end{equation}
of the saddle-point solution
\begin{equation}
  \Lambda_{tt'}
  =
  \begin{pmatrix}
    \delta^R_{t t'} & 0 \\
    0 & -\delta^A_{t t'}
  \end{pmatrix}
\label{lambda-def}
\end{equation}
under unitary rotations $U$. The integration measure $[DQ]$ is then
the usual invariant measure on the orbit.

We want to stress that the present approach slightly differs from
the scheme originally proposed in Refs.~\onlinecite{skvor,BSK03,SBK04}.
The sigma-model derived in Ref.~\onlinecite{skvor}, having the same
action (\ref{action-general}), was formulated on a different manifold
$Q_F=U_F^{-1}U^{-1}\Lambda UU_F=U_F^{-1}QU_F$, where $U_F$ is a non-Hermitian
rotation which contains the knowledge about the fermionic distribution
function [see Eq.~(\ref{UF}) for an explicit form of $U_F$].
However, since calculating the energy-space diffusion coefficient $D$
is essentially a {\em single-particle}\/ problem,
one can get rid of the distribution function $F$ in the definition
of the integration manifold and express the diffusion coefficient $D$
in terms of a certain correlation function of the fields $Q$.
This refinement of the theory is presented in Appendix \ref{A:Keldysh}.

The quantity of our interest will be the {\it diffuson} ${\cal D}_{\eta}(t)$
defined by
\begin{equation}
\langle  Q^{(+)}_{t_1,t_2} \, Q^{(-)}_{t_3,t_4} \rangle = \int
[DQ]\, e^{-S[Q]} Q^{(+)}_{t_1,t_2} \, Q^{(-)}_{t_3,t_4} =
\frac{2}{\pi}
 \delta(t_1-t_2+t_3-t_4)
{\cal D}_{t_1-t_2}(t_1-t_4)\, ,
\label{full-diffuson-definition}
\end{equation}
where
\begin{equation}
  Q^{(\pm)}_{t_1 t_2}=\Tr_K (\sigma^\mp Q_{t_1 t_2})
\label{Q-off-diag}
\end{equation}
 are the
off-diagonal elements of the $Q$-matrix in the Keldysh space.
The form of the right-hand side in (\ref{full-diffuson-definition})
follows from the invariance of the action $S[Q]$
with respect to time translations ($Q_{tt'}\mapsto Q_{t+\delta
t,t'+\delta t}$) and with respect to the energy shift
($Q_{tt'}\mapsto Q_{tt'}e^{i\omega(t-t')}$). Using the causality
principles, one can further show (see Section
\ref{section-diagrams} and Appendix \ref{A:Keldysh})
that the diffuson must have the form
\begin{equation}
{\cal D}_\eta(t)= \theta(t) \exp [-P(\eta,t)],
\label{full-diffuson}
\end{equation}
where $P(\eta=0,t)=0$. Further we may expand $P(\eta,t)$ in $\eta$.
The diffusion coefficient $D(\alpha)$ defined in (\ref{diffusion-definition})
is given by the coefficient at $P(\eta,t)$ at $\eta^2 t$ (at small $\eta$
and at large $t$). The derivation is given in Appendix~\ref{A:Keldysh}.
Formally we may write
\begin{equation}
D(\alpha)= - \lim_{t\to\infty} \frac{1}{2t}
\left. \frac{\partial^2}{\partial \eta^2} \right|_{\eta=0}
{\cal D}_\eta(t) .
\label{diffusion-coefficient}
\end{equation}

\section{Diagrammatic expansion}
\label{section-diagrams}

To develop the diagrammatic technique, we explicitly parameterize the
unitary rotations $U$ in (\ref{orbit}) by elements of the
corresponding Lie algebra. This parameterization may be chosen in
many different ways, and we write generally
\begin{equation}
U=f^{1/2}(W)\, ,
\label{parameter-general-1}
\end{equation}
where the components of $W$ are given by
\begin{equation}
  W_{tt'}
  =
  \begin{pmatrix}
    0 & b_{tt'} \\
    -\bar{b}_{tt'} & 0
  \end{pmatrix} ,
\label{W-b}
\end{equation}
and $f(W)$ may be represented as the series
\be
  f(W)=1+W+W^2/2 + c_3 W^3 + c_4 W^4 + \dots
\label{f-series}
\ee
The unitarity of $U$ implies $\bar{b}_{tt'} = b^*_{t't}$ and $f(W)f(-W)=1$.
We have chosen to write $f^{1/2}$ in the
definition (\ref{parameter-general-1}) so that the
parameterization of $Q$ contains the first power of $f$:
\begin{equation}
  Q = \Lambda f(W)
\label{parameter-general-2}
\end{equation}
(note that $W$ anticommutes with $\Lambda$).

Possible choices of the parameterization will
be discussed in detail in Appendix \ref{A:parameterizations}.
Using the parameterization (\ref{W-b}), (\ref{parameter-general-2}),
the integration measure $[DQ]$
may be written as $[Db\, D\bar{b}] J_f[b,\bar{b}]$, where
$J_f$ is the Jacobian associated with the parameterization $f$.
The explicit expression for the Jacobian in terms of the
function $f$ is given in Appendix \ref{A:parameterizations}.

Then we do the algebra of substituting the parameterization
(\ref{lambda-def}), (\ref{f-series}), (\ref{parameter-general-2})
into the action (\ref{action-total})--(\ref{action-kin}) to obtain
\begin{equation}
S[b,\bar{b}]=S^{(2)}[b,\bar{b}]+ S^{(\ge 4)}_{\rm E} [b,\bar{b}]+
S^{(\ge 4)}_{\rm kin} [b,\bar{b}],
\label{S[b]}
\end{equation}
where
\begin{equation}
S^{(2)}[b,\bar{b}] = \frac{\pi}{2} \int\!\!\int dt_1\, dt_2\,
\bar{b}_{12}\Big[ (\partial_1 + \partial_2) + \alpha(t_1-t_2)^2 \Big]
b_{21}
\label{b-action-quadratic}
\end{equation}
is the quadratic part of the action, and  $S^{(\ge 4)}_{\rm E} [b,\bar{b}]$
and $S^{(\ge 4)}_{\rm kin} [b,\bar{b}]$ are the higher-order terms
(only even-order terms in $b$, $\bar{b}$ appear in the action).

The propagator of the quadratic action $S^{(2)}[b,\bar{b}]$ is the
{\it bare diffuson}:
\be
  \langle b_{t+\eta/2,t-\eta/2} \bar{b}_{t'-\eta'/2,t'+\eta'/2} \rangle^{(0)}
  = \frac{2}{\pi}
  \delta(\eta-\eta')
  {\cal D}^{(0)}_{\eta}(t-t')
\label{bare-propagator}
\ee
with
\begin{equation}
{\cal D}^{(0)}_\eta(t)=\theta(t) \exp[-\alpha\eta^2 t] .
\label{bare-diffuson}
\end{equation}

\begin{figure}
\epsfxsize=0.5\hsize
\centerline{\epsfbox{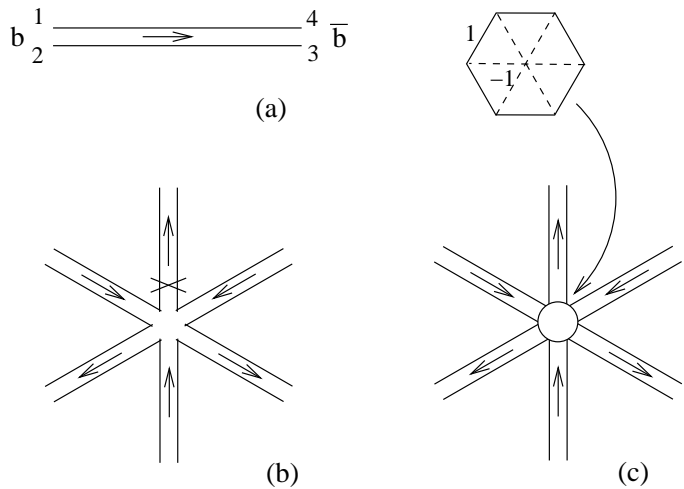}}
\medskip
\caption{
{\bf (a)} Bare diffuson ${\cal D}^{(0)}_\eta(t)$. The arrow
points in the direction of decreasing time. {\bf (b)} Vertices arising from
$S^{(\ge 4)}_{\rm E} [b,\bar{b}]$. Crossed diffuson denotes differentiation
$\hat\partial b$ in the vertex (\ref{E-vertex}). {\bf (c)} Vertices
arising from $S^{(\ge 4)}_{\rm kin} [b,\bar{b}]$. The vertex itself is
characterized by a graph connecting different
times $t_i$ at this vertex. The numbers on the links of the graph specify
the coefficients $a^{(n)}_{ij}$ in (\ref{kin-vertex}).
}
\label{fig:diagram-elements}
\end{figure}

Now we build up the diagrammatic expansion: by expanding
the higher-order terms $S^{(\ge 4)}_{\rm E} [b,\bar{b}]$
and $S^{(\ge 4)}_{\rm kin} [b,\bar{b}]$ we generate vertices which we
then connect by the propagators (\ref{bare-propagator}) of the
quadratic theory. Each propagator (\ref{bare-propagator}) will be
graphically represented as a band with two incoming legs (corresponding
to the $b$-end of the diffuson) and two outgoing legs (the $\bar{b}$-end
of the diffuson), see Fig.~\ref{fig:diagram-elements}. These two ends are
not equivalent: the $b$-vertex must contain later times than the $\bar{b}$
vertex (due to the retarded $\theta(t)$ factor in the diffuson
(\ref{bare-diffuson})). Correspondingly, we will draw an arrow
on the diffuson line pointing in the direction of decreasing time,
from $b$ to $\bar{b}$.

The nonlinear vertices arising from $S^{(\ge 4)}_{\rm E} [b,\bar{b}]$
have the form
\begin{equation}
  S^{(2n)}_{\rm E} [b,\bar{b}]
  =
  (-1)^{n+1} \pi \, c_{2n}
  \Tr [(\hat\partial b)\bar{b} (b\bar{b})^{n-1}] ,
\label{E-vertex}
\end{equation}
where we use the shorthand notation $(\hat\partial b)_{12}=
(\partial_1+\partial_2)b_{12}$, and the trace is understood as
the convolution in time variables. The vertices from
$S^{(\ge 4)}_{\rm kin} [b,\bar{b}]$ are of the form
\begin{equation}
  S^{(2n)}_{\rm kin} [b,\bar{b}]
  =
  \alpha
  \int dt_1 \dots dt_{2n} \,
  b_{12} \bar{b}_{23} \dots \bar{b}_{2n,1} \sum_{i<j} a_{ij}^{(n)}(t_i-t_j)^2 ,
\label{kin-vertex}
\end{equation}
where
$a_{ij}^{(n)}$ are numerical coefficients expressed
through the coefficients $c_k$ of the expansion (\ref{f-series}):
\be
  a_{ij}^{(n)} = \frac{\pi}{2n} (-1)^{i-j+n} c_{j-i} c_{i-j+2n} .
\ee
Graphically, we order the diffusons at each vertex
clockwise according to their appearance in the traces (\ref{E-vertex}) and
(\ref{kin-vertex}).

The average value of the physical observables is given by
certain correlators of the original $Q$-field.
When we compute them in terms of the $b$ and $\bar{b}$ fields,
those observables generate ``external vertices''.
The vertices generated by
$S^{(\ge 4)}_{\rm E} [b,\bar{b}]$ and $S^{(\ge 4)}_{\rm kin} [b,\bar{b}]$
will be further called ``internal vertices''. Finally,
vertices generated by the Jacobian $J_f[b,\bar{b}]$ (unless it equals one)
we shall call ``Jacobian vertices'' (see Appendix \ref{A:parameterizations}
for the explicit form of the Jacobian vertices).

\begin{figure}
\epsfxsize=0.4\hsize
\centerline{\epsfbox{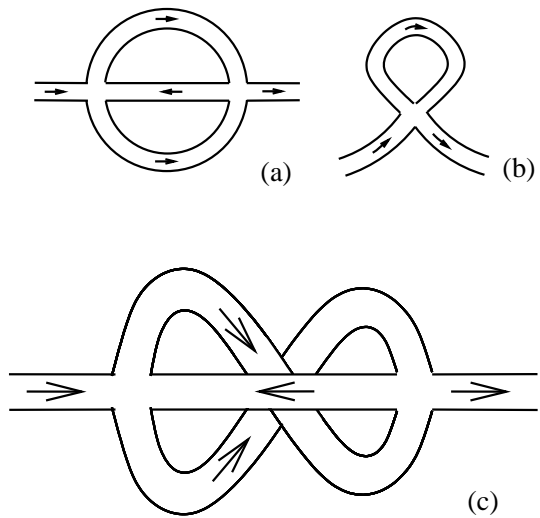}}
\medskip
\caption{
Examples of diagrams. Diagrams {\bf (a)} and {\bf (b)} trivially vanish.
Panel {\bf (c)} shows a non-vanishing two-loop diagram (this diagram is one
of those considered in Refs.~\onlinecite{skvor,SBK04}).
}
\label{fig:diagram-examples}
\end{figure}

There are two obvious rules for constructing the diagrams. Firstly,
the diffusons may be drawn on a planar figure without ``twisting''
(but intersections of different diffusons are allowed, see, e.g.,
Fig.~\ref{fig:diagram-examples}c). Secondly, the diagram must not
contain closed loops formed by internal lines. Such loops would
immediately produce the factor $\theta(t_1-t_2)\theta(t_2-t_3)\dots
\theta(t_{n}-t_1)$ which always vanishes [note that $\theta(t_1-t_1)=0$
due to the causality rule~\cite{KamenevAndreev99,AltlandKamenev00},
and so loops of length one
are not allowed either]. For example, the diagrams in
Figs.~\ref{fig:diagram-examples}ab identically vanish and should not be
considered. An example of a non-vanishing two-loop diagram is shown
in Fig.~\ref{fig:diagram-examples}c.

{} From the above constraints on the diagram we can prove that only
even-loop diagrams contribute to the diffuson (\ref{full-diffuson}).
More generally, the following {\it theorem} holds:
Let $b^{(2n+1)}$ and $\bar{b}^{(2n+1)}$ denote the operators
\begin{eqnarray}
b^{(2n+1)}_{tt'}&=&
\int b_{t t_1} \bar{b}_{t_1 t_2} b_{t_2 t_3} \dots b_{t_{2n} t'}
\, dt_1\,\dots\, dt_{2n} ,
\nonumber \\
\bar{b}^{(2n+1)}_{tt'}&=&
\int \bar{b}_{t t_1} b_{t_1 t_2} \bar{b}_{t_2 t_3} \dots \bar{b}_{t_{2n} t'}
\, dt_1\,\dots\, dt_{2n} .
\label{b(n)}
\end{eqnarray}
Then any nonzero average
$\corr{b^{(n_1)}\dots b^{(n_k)}\, \bar{b}^{(\bar{n}_1)}
\dots\bar{b}^{(\bar{n}_{\bar{k}})}}$ must contain equal number of $b$- and
$\bar{b}$-operators and its diagrammatic expansion contains only diagrams
with {\it even}\/ number of loops.

Furthermore, from the same principle it is easy to prove that
$\langle b^{(n)}_{t_1 t_2} \bar{b}^{(\bar{n})}_{t_3 t_4}\rangle=0$ if
$t_1\le t_4$ in all orders of the diagrammatic expansion.
This proves the causality of the full diffuson
(\ref{full-diffuson-definition}): ${\cal D}_\eta(t\le 0) =0$.

In the main body of the paper,
we shall only use the {\it rational}\/ parameterization:
\begin{equation}
  f(W) = \frac{1 + W/2}{1 - W/2} .
\label{parameter-rational}
\end{equation}
The Jacobian for the rational parameterization
is known~\cite{Efetov1983,Efetov-book} to be equal to 1.
(In fact, such a parameterization is not unique.
As shown in Appendix \ref{A:parameterizations},
the class of parameterizations with unit Jacobian
is given by a one-parameter family, with the rational
parameterization being a particular representative.)

In the rational parameterization (\ref{parameter-rational}),
the coefficients of the series (\ref{f-series})
are given by $c_k=2^{1-k}$ ($k\geq1$)
and the coefficients $a_{ij}^{(n)}$ in the nonlinear
vertices (\ref{kin-vertex}) have the form
\be
  a_{ij}^{(n)}
  = \frac{1}{n} \, (-1)^{i-j+n} 2^{1-2n}\pi .
\label{a-rational}
\ee

\section{Diagrammatic series for the diffusion coefficient}
\label{section-diffuson}

In this section we apply the developed diagrammatic technique to calculating
the full diffuson (\ref{full-diffuson-definition}) and to further
extracting the diffusion coefficient (\ref{diffusion-coefficient}).

The observables in (\ref{full-diffuson-definition}) have the form
\begin{eqnarray}
Q^{(+)} &=& b^{(1)} - c_3 b^{(3)} + c_5 b^{(5)} - \dots , \cr
Q^{(-)} &=& \bar{b}^{(1)} - c_3 \bar{b}^{(3)} + c_5 \bar{b}^{(5)} - \dots ,
\end{eqnarray}
where $c_3$, $c_5$, \dots are the coefficients in the Taylor
expansion of the parameterization (\ref{f-series}), and the
operators $b^{(n)}$ and $\bar b^{(n)}$ are defined in Eq.~(\ref{b(n)}).
Respectively, the average $\langle Q^{(+)} Q^{(-)} \rangle$ contains
averages of different powers of $b$ and $\bar{b}$.

\begin{figure}
\centerline{\epsfbox{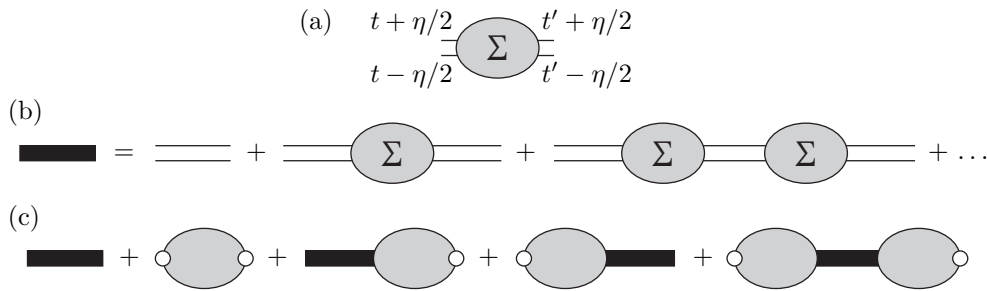}}
\medskip
\caption{
(a) Irreducible self-energy block $\Sigma_\eta(t-t')$.
(b) The diagrammatic expansion of the pro-diffuson $\langle b \bar{b} \rangle$
denoted by the solid propagator.
(c) The diagrammatic expansion of the full diffuson
$\langle Q^{(+)} Q^{(-)} \rangle $.
The irreducible blocks differ from $\Sigma$
due to the presence of {\it external} vertices (shown by empty circles)
containing higher powers of $b$ and $\bar{b}$.}
\label{fig:diffuson-diagram}
\end{figure}

Let us first discuss the simplest of those averages:
\be
  \langle b_{t+\eta/2,t-\eta/2} \bar{b}_{t'-\eta'/2,t'+\eta'/2} \rangle
  = \frac{2}{\pi}
  \delta(\eta-\eta')
  {\cal D}^*_{\eta}(t-t') .
\label{D-star-propagator}
\ee
Eq.~(\ref{D-star-propagator}) defines the ``pro-diffuson'' ${\cal D}^*_\eta(t)$
which should not be confused with the full diffuson
given by Eq.~(\ref{full-diffuson-definition}).
The pro-diffuson ${\cal D}^*_\eta(t)$ may be written
in the perturbative series as shown in
Fig.~\ref{fig:diffuson-diagram}b. Defining the irreducible
``self-energy'' block $\Sigma_\eta(t)$
(shown pictorially in Fig.~\ref{fig:diffuson-diagram}a)
by the condition that it cannot be split in two pieces
by cutting any one diffuson, we
can resum the diagrammatic series for the pro-diffuson as
\be
  \bigl[ {\cal D}^*_\eta(\omega) \bigr]^{-1}
  =
  \bigl[ {\cal D}^{(0)}_\eta(\omega) \bigr]^{-1}
  - \Sigma_\eta(\omega) .
\label{D-star-Sigma}
\ee
Here we adopted the frequency representation, with
\be
  {\cal D}^{(0)}_\eta(\omega)
  = \int_{-\infty}^{+\infty}
    e^{i\omega t} {\cal D}^{(0)}_\eta(\omega)\, dt
  = \frac{1}{-i\omega+\alpha\eta^2} .
\label{D0-freq}
\ee
The role of the self-energy $\Sigma_\eta(\omega)$
is shifting the pole of the pro-diffuson ${\cal D}^*_\eta(\omega)$.

One can see that the bare diffuson (\ref{D0-freq}) in the frequency
representation looks exactly as the diffuson propagator
$1/(-i\omega+Dq^2)$ for a particle in a disordered medium.
However, the analogy between these problems does not extend
beyond the formal coincidence of the bare propagators.
Indeed, for a particle in a random potential,
the frequency $\omega$ and momentum $q$ are ``decoupled''
from each other: each diffuson in a diagram has the same
frequency, while the choice of their momenta $q_i$
is dictated by the momentum conservation law in the vertices.
Contrary, for the dynamic problem, the times $\eta_i$ and $t_j$
of the internal diffusons ${\cal D}^{(0)}_{\eta_i}(t_i)$ which
constitute the self-energy block $\Sigma_\eta(t)$ are linearly related
to each other (see Section \ref{section-diagrams-D} for their
explicit form).
Therefore, in the frequency representation,
the internal diffusons in $\Sigma_\eta(\omega)$ will enter
with different frequencies $\omega_i$ coupled
to the time arguments $\eta_j$.
Thus, the frequency representation (\ref{D0-freq}) is not a way
to calculate $\Sigma_{\eta}(t)$, but is a convenient tool
for summing the geometric series (\ref{D-star-Sigma}).

Note that the full diffuson ${\cal D}_\eta(t)$ differs from the pro-diffuson
${\cal D}^*_\eta(t)$ by higher-power averages
$\langle b^{(n)} \cdot \bar{b}^{(\bar{n})} \rangle$
(see Fig.~\ref{fig:diffuson-diagram}c). We observe that
{\it all}\/ irreducible parts [both $\Sigma_\eta(t)$ and the
irreducible blocks in Fig.~\ref{fig:diffuson-diagram}c] remain
exponentially decaying with time even at $\eta=0$ (in the frequency
representation,
they are non-singular functions of $\eta$ and $\omega$ at $\eta \to 0$
and $\omega\to 0$)~\cite{convergence}.
Therefore we may sum all the diagrams to the form
\begin{equation}
{\cal D}_\eta(\omega)=\frac{Z(\eta,\omega)}{-i\omega + \alpha\eta^2
- \Sigma_\eta(\omega)} ,
\end{equation}
where $Z(\eta,\omega)$ is a regular function at $\eta \to 0$
and $\omega\to 0$. As we shall see from the further explicit
calculations,
$\Sigma_\eta(\omega)\to 0$ as $\eta\to 0$.
Furthermore, the full diffuson ${\cal D}_\eta(\omega)$
remains unrenormalized at $\eta=0$
(see Appendix~\ref{A:Keldysh} for a proof).
Therefore, $Z(\eta=0,\omega)=1$.

Thus the correction to the diffusion coefficient is determined by the
leading term in the expansion of $\Sigma_\eta(\omega)$ in $\eta$
and $\omega$:
\be
  D(\alpha)
  = \alpha
  - \lim_{\eta\to 0} \frac{\Sigma_\eta(\omega=0)}{\eta^2}
  = \alpha
  - \lim_{\eta\to 0} \frac{1}{\eta^2} \int_0^\infty \Sigma_\eta(t) \, dt ,
\label{D-correction}
\ee
where the lower limit of $t$-integration
follows from the causality of $\Sigma_\eta(t)$ which can be proved
analogously to the causality of the diffuson ${\cal D}_\eta(t)$.

The diagrams representing the irreducible block $\Sigma_\eta(t)$
may be classified in the number of loops $L$.
From simple power counting (as shown in
Ref.~\onlinecite{skvor}) the $L$-loop diagrams give a contribution to
$D(\alpha)$ proportional to $\alpha^{1-L/3}$, cf.\ Eq.~(\ref{D-series}).
Therefore, the diagrammatic expansion is a series in the number of loops,
and (since the number of loops $L$ must be even) the expansion small
parameter is $\alpha^{-2/3}$, up to some unknown number.
A more accurate analysis of Sec.~\ref{section-diagrams-D} shows that
the actual small parameter is $\pi^{-2}\alpha^{-2/3}$.

In the following section we shall see that certain different diagrams
with the same number of loops $L$ may partially cancel each other.

\section{Canceling vertices of order higher than four}
\label{section-cancellation}

In this section we show that, in the {\it rational}\/
parameterization (\ref{parameter-rational}), in diagrams
containing only internal vertices (e.g., in $\Sigma_\eta(t)$),
{\it all}\/ vertices of order higher than four are cancelled
{\it in all orders}\/ of the diagram series.

The outline of the calculation is as follows. The vertices with
derivatives originating from $S_{\rm E}^{(\geq4)}$
(Fig.~\ref{fig:diagram-elements}b) generate terms
$\partial_t {\cal D}^{(0)}_\eta (t)$. We transform those terms
according to the {\it equation of motion} for the diffuson:
\begin{equation}
\partial_t {\cal D}^{(0)}_\eta (t) = \delta(t) - \alpha \eta^2
{\cal D}^{(0)}_\eta (t) .
\label{eq-motion}
\end{equation}
The resulting expressions have the form of $(t_i-t_j)^2$
vertices in Fig.~\ref{fig:diagram-elements}c and may then be
combined with the vertices originally generated by
$S^{(\ge 4)}_{\rm kin} [b,\bar{b}]$.

When performed in the rational parameterization (\ref{parameter-rational}),
this procedure leads to the cancellation of all vertices of order
higher than four in all orders of the perturbation series. The calculation
is somewhat technical with the combinatoric counting of diagrams
and coefficients. This calculation is reported in detail in Appendix \ref{A:rational}.

Of course, at the end of the above simplification procedure, the
fourth-order vertices get modified from their original form.
The final form of the fourth-order vertex is
\be
  S^*_4[b,\bar{b}]
  = - \frac{\pi\alpha}{8} \int dt_1\, dt_2\, dt_3\, dt_4 \,
  (t_1-t_2)(t_3-t_4) \,
  b_{12}\bar{b}_{23}b_{34}\bar{b}_{41} .
\label{S-star}
\ee
The resulting theory in terms of $b$-fields
(which we will further call ``$b$-theory'') has the action
\be
  S_{\rm eff}[b,\bar b]
  =
  S^{(2)}[b,\bar{b}] + S^*_4[b,\bar{b}] ,
\label{b-action-rational}
\ee
where $S^{(2)}[b,\bar{b}]$ is given by (\ref{b-action-quadratic}).

We need to make two comments on the above derivation.

First, up to now
the correspondence is established only between diagrams containing
{\it internal} vertices. The correspondence between
{\it physical observables} (i.e., external vertices)
will be discussed in Section \ref{section-DM}.
In the further discussion, we employ the $b$-theory
for calculating $\Sigma_\eta(t)$ which contains only
internal vertices.

Second, when constructing the diagrams for  $\Sigma_\eta(t)$
in the $b$-theory, one of the $(t_{2i-1}-t_{2i})$ factors in $S^*_4[b,\bar{b}]$
always coincides with the external time difference $\eta$. The
integrals over times $t_i$ converge~\cite{convergence} and therefore
\begin{equation}
  \Sigma_{\eta=0}(t)=0 .
\end{equation}
Obviously $\Sigma_\eta(t)$ is an even and regular function of $\eta$.
Thus the Taylor expansion of $\Sigma_\eta(t)$ in small $\eta$ starts
with $\eta^2$. The leading term in this expansion determines the correction
to the diffusion coefficient, according to (\ref{D-correction}). We shall
perform an explicit calculation in the next section.

\section{Diagrammatic expansion for $D(\alpha)$}
\label{section-diagrams-D}

In this section we classify the diagrams for $\Sigma_\eta(t)$ in
the $b$-theory (\ref{b-action-rational}) and write down
explicit integral expressions
for the correction to the diffusion coefficient (\ref{D-correction}).

\begin{figure}
\epsfxsize=0.4\hsize
\centerline{\epsfbox{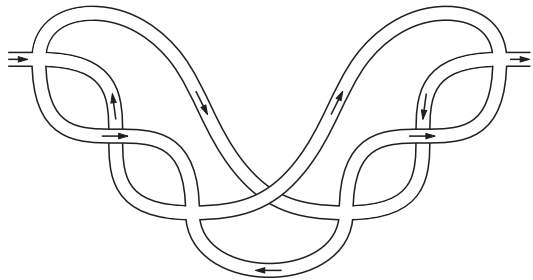}}
\caption{An example of the 4-loop diagram for the self-energy $\Sigma$.
This diagram corresponds to the permutation $\sigma=[3764215]$
listed as No.\ 2 in the Table \ref{table:four-loop}.}
\label{fig:loop4example}
\end{figure}

Consider any diagram (in the $b$-theory) with $L$ loops
contributing to $\Sigma_\eta(t)$. As mentioned
previously, such a diagram must contain no closed threads, which greatly
restricts the number of allowed diagrams (in particular, the number of
loops $L$ must be even). Since all the vertices in the $b$-theory
have valency four, the number of diffusons in the diagram is $2L-1$. We
enumerate those diffusons in two different sequences which we shall
call the ``left-hand'' and ``right-hand'' numberings. Start at the
``entry point'' (external diffuson leading into the diagram) and go along
the diffusons in two different ways: using the right-hand rule (following
the right edges of the diffusons, i.e., turning right at every vertex)
and using the left-hand rule (following
the left edges of the diffusons and turning left at every vertex).
The relative re-numbering of the diffusons in the two sequences defines
a permutation $\sigma$ of $(2L-1)$ elements. In other words, let us number
the diffusons following the {\it right-hand} route. Then reading off the
diffuson numbers along the {\it left-hand} route produces the sequence of
numbers [$\sigma(1)$, $\sigma(2)$,\dots, $\sigma(2L-1)$]. We shall further
label the diagrams by such sequences (see Table~\ref{table:four-loop}).
An example of the 4-loop diagram corresponding to the permutation
$\sigma=[3764215]$ (numbered 2 in the Table~\ref{table:four-loop})
is shown in Fig.~\ref{fig:loop4example}.

\begin{table}
\tabcolsep=2mm
\begin{tabular}{|c|c|l|}
\hline
No. & $\sigma$ & comments \\
\hline
\hline
1 & [3214765] & reducible \\
\bf2 & [3764215] & \\
\bf3 & [3752164] & \\
\bf4 & [3621754] & \\
\bf5 & [3654721] & \\
6 & [4276315] & = No.\ 3 (T) \\
\bf7 & [4731652] & \\
8 & [4317625] & = No.\ 4 (LR/T) \\
\bf9 & [4726531] & \\
\bf10& [5274163] & \\
11& [5417362] & = No.\ 3 (LR) \\
\bf12& [5372641] & \\
13& [6327514] & = No.\ 7 (T) \\
14& [6514732] & = No.\ 2 (LR) \\
15& [6251743] & = No.\ 3 (LR+T) \\
16& [6437251] & = No.\ 9 (LR+T) \\
17& [7614325] & = No.\ 5 (LR/T) \\
18& [7532614] & = No.\ 9 (T) \\
19& [7426153] & = No.\ 12 (LR/T) \\
20& [7361542] & = No.\ 9 (LR) \\
\bf21& [7254361] & \\
\hline
\end{tabular}
\caption{The list of all four-loop diagrams for the self-energy part $\Sigma$.
Each diagram is characterized by a permutation $\sigma$, see text.
The diagram No.~1 is reducible: it can be written
as $\Sigma_2 {\cal D}^{(0)} \Sigma_2$, where $\Sigma_2$ is the two-loop
contribution. The other diagrams are irreducible.
The diagrams related by the ``left-right'' (LR) or time-reversal (T)
symmetries give the same contribution.
LR+T means that two diagrams are related by the combined action of the
two symmetries. LR/T indicate that two symmetries applied to
a permutation give the same result.
Graphical representation of the diagram No.\ 2
is given in Fig.~\ref{fig:loop4example}.}
\label{table:four-loop}
\end{table}

Note that not every permutation of $2L-1$ elements defines a diagram.
We do not discuss here the combinatoric problem of enumerating all diagrams
in all orders. For our modest purpose of calculating the four-loop correction
to the diffusion coefficient,
the enumeration of the diagrams may be easily done
``by hand''. Recall that for $L=2$ there is only one allowed diagram
(shown in Fig.\ \ref{fig:diagram-examples}) corresponding to the permutation
[3,2,1]. For $L=4$, the number of allowed diagrams (and the corresponding
permutations) is 21, see Table~\ref{table:four-loop}. Note that one of those
diagrams (No.\ 1 in Table~\ref{table:four-loop})
is {\it reducible}: it consists of two independent two-loop blocks.

Given a diagram characterized by a certain permutation $\sigma$,
let $\eta_i$ and $t_i$ denote
the parameters of the diffusons ${\cal D}^{(0)}_{\eta_i} (t_i)$ involved
in the diagram, with the diffusons enumerated along the right-hand route.
For calculating the diagram, it is convenient to take
$t_1,\dots,t_{2L-1}$ as independent integration
variables. The parameters $\eta_i$ may be expressed via $t_i$ as
\begin{equation}
  \eta_i
  = \eta + \sum_{j<i} t_j - \sum_{\sigma(j)<\sigma(i)} \!\!\!\! t_j
  = \eta + \sum_j \gamma^{(\sigma)}_{ij} t_j .
\label{eta-i}
\end{equation}
Here the first (second) sum contains all times $t_j$ encountered before
$t_i$ along the right-hand (left-hand, respectively) route.
The last equality is the definition of the coefficients $\gamma^{(\sigma)}_{ij}$.
The matrix $\gamma^{(\sigma)}_{ij}$ contains only entries
$0$, $1$, and $-1$, and is antisymmetric: $\gamma^{(\sigma)}_{ij}=-\gamma^{(\sigma)}_{ji}$.
Moreover, the elements $\gamma^{(\sigma)}_{ij}$ with $i>j$ are non-negative (either 0 or 1),
and those with $i<j$ are non-positive (either 0 or $-1$).

Finally, calculating the self-energy block $\Sigma_\eta(t)$
in the $b$-theory (\ref{b-action-rational}) and employing
Eq.~(\ref{D-correction}), we may write down
the contribution of any given irreducible diagram (corresponding to a
given permutation $\sigma$) to the diffusion coefficient $D(\alpha)$:
\begin{equation}
  \delta^{(\sigma)} D(\alpha)
  = - \frac{K^{(\sigma)}}{\pi^L} \alpha^{1-L/3} ,
\end{equation}
where
\be
  K^{(\sigma)}
  = \int_0^\infty \!\!\!\!\dots \int_0^\infty
    \left( \prod_{i=1}^{2L-1} dT_i \right)
    P^{(\sigma)} (T_1, \dots, T_{2L-1})
    e^{-S_3^{(\sigma)}(T_1, \dots, T_{2L-1})} .
\label{K-sigma}
\ee
Here $P^{(\sigma)}(T_1, \dots, T_{2L-1})$
and $S_3^{(\sigma)}(T_1, \dots, T_{2L-1})$ are
homogeneous polynomials of degrees $2L-2$ and three, respectively.
They are defined as
\begin{subequations}
\begin{align}
  P^{(\sigma)}(T_1, \dots, T_{2L-1}) &=
  \left(\prod_{i=1}^{2L-1} {\tilde\eta}_i \right)
  \sum_{i=1}^{2L-1} \frac{1}{{\tilde\eta}_i} ,
\\
  S_3^{(\sigma)}(T_1, \dots, T_{2L-1}) &=
  \sum_{i=1}^{2L-1} {\tilde\eta}_i^2 T_i ,
\\
  {\tilde\eta}_i &=
  \sum_j \gamma^{(\sigma)}_{ij} T_j ,
\end{align}
\end{subequations}
where the coefficients $\gamma^{(\sigma)}_{ij}$
are constructed from the permutation
$\sigma$ according to Eq.~(\ref{eta-i}).

\begin{table}
\tabcolsep=2mm
\begin{tabular}{|c|c|r|r|}
\hline
No. & $M^{(\sigma)}$ & $K^{(\sigma)}$ & st. dev.\\
\hline
\hline
\bf2 & 2 & 0.066945    & 0.000015 \\ 
\bf3 & 4 & $-0.123205$ & 0.000012 \\ 
\bf4 & 2 & 0.082358    & 0.000018 \\ 
\bf5 & 2 & 0.008234    & 0.000005 \\ 
\bf7 & 2 & 0.006030    & 0.000007 \\ 
\bf9 & 4 & $-0.006567$ & 0.000005 \\ 
\bf10& 1 & 0.206687    & 0.000013 \\ 
\bf12& 2 & $-0.001646$ & 0.000006 \\ 
\bf21& 1 & $-0.011374$ & 0.000003 \\ 
\hline
\end{tabular}
\caption{Results of numerical evaluation of the diagrams
from Table~\ref{table:four-loop}. $M^{(\sigma)}$ denote
the multiplicities of the diagrams. $K^{(\sigma)}$ is given
by Eq.~(\ref{K-sigma}). The last column is the standard
deviation of $K^{(\sigma)}$.}
\label{table:numerics}
\end{table}

\section{Numerical calculation of the 4-loop diagrams}
\label{section-four-loops}

The list of allowed four-loop diagrams (permutations) is given
in Table \ref{table:four-loop}.
The first diagram in the list is not irreducible: it consists
of two blocks of two loops each and should not be included in
the irreducible part $\Sigma_\eta(t)$. Out of the remaining 20
diagrams, some are related by symmetries and produce the same
contributions (\ref{K-sigma}). Namely, there are
two possible symmetries: the ``left-right'' (LR) reflection of
the diffusons (corresponding to the transformation
$\sigma \mapsto \sigma^{-1}$) and the time-reversal (T) symmetry
(corresponding to replacing $\sigma(i)\mapsto 2L-\sigma(2L-i)$).
The diagrams related by the symmetries are identified in
the above table. It remains to calculate the nine different
diagrams and add them with the corresponding multiplicities
$M^{(\sigma)}$, see Table \ref{table:numerics}.

The antisymmetric matrices $\gamma^{(\sigma)}$ defined for nine permutations $\sigma$
from Table \ref{table:numerics} are listed below, with the elements $+$ ($-$)
standing for 1 ($-1$) for brevity:

\begin{eqnarray}
\gamma^{(2)} = \left[ \begin {array}{ccccccc} 0&-&-&-&0&-&-\\+&0&
-&-&0&-&-\\+&+&0&0&0&0&0\\+&+&0
&0&0&-&-\\0&0&0&0&0&-&-\\+&+&0&+&+&0&-\\+&+&0&+&+&+&0\end {array} \right];
\quad
\gamma^{(3)} = \left[ \begin {array}{ccccccc} 0&-&-&0&-&0&-\\+&0&
-&0&-&0&-\\+&+&0&0&0&0&0\\0&0&0&0
&-&-&-\\+&+&0&+&0&0&-\\0&0&0&+
&0&0&-\\+&+&0&+&+&+&0\end {array} \right];
\quad
\gamma^{(4)} = \left[ \begin {array}{ccccccc} 0&-&-&0&0&-&0\\+&0&
-&0&0&-&0\\+&+&0&0&0&0&0\\0&0&0&0
&-&-&-\\0&0&0&+&0&-&-\\+&+&0&+
&+&0&0\\0&0&0&+&+&0&0\end {array} \right];
\nonumber \\
\gamma^{(5)} = \left[ \begin {array}{ccccccc}
0&-&-&-&-&-&-\\
+&0&-&-&-&-&-\\
+&+&0&0&0&0&0\\
+&+&0&0&-&-&0\\
+&+&0&+&0&-&0\\
+&+&0&+&+&0&0\\
+&+&0&0&0&0&0
\end {array} \right];
\quad
\gamma^{(7)} = \left[ \begin {array}{ccccccc} 0&0&-&-&0&0&-\\0&0&-
&-&-&-&-\\+&+&0&-&0&0&-\\+&+&+&0&0&0&0\\0&+&0&0&0&-&-\\0&+&0&0
&+&0&-\\+&+&+&0&+&+&0\end {array} \right];
\quad
\gamma^{(9)} = \left[ \begin {array}{ccccccc} 0&-&-&-&-&-&-\\+&0&0
&-&0&0&-\\+&0&0&-&-&-&-\\+&+&+
&0&0&0&0\\+&0&+&0&0&-&-\\+&0&+
&0&+&0&-\\+&+&+&0&+&+&0\end {array} \right];
\\
\gamma^{(10)} = \left[ \begin {array}{ccccccc} 0&-&0&-&-&0&-\\+&0&0
&0&-&0&0\\0&0&0&-&-&-&-\\+&0&+&0
&-&0&-\\+&+&+&+&0&0&0\\0&0&+&0
&0&0&-\\+&0&+&+&0&+&0\end {array} \right];
\quad
\gamma^{(12)} = \left[ \begin {array}{ccccccc} 0 & - & - & - & - & - & - \\
    + & 0 & - & 0 & - & 0 & - \\ + &
    + & 0 & 0 & - & 0 & 0 \\
    + & 0 & 0 & 0 & - & - & - \\ + & + & + &
    + & 0 & 0 & 0 \\ + & 0 & 0 &
    + & 0 & 0 & - \\ + & + & 0 & + & 0 &
    + & 0 \end {array} \right] ;
\quad
\gamma^{(21)} = \left[ \begin {array}{ccccccc} 0&-&-&-&-&-&-\\+&0&0
&0&0&0&-\\+&0&0&-&-&0&-\\+&0&+&0
&-&0&-\\+&0&+&+&0&0&-\\+&0&0&0
&0&0&-\\+&+&+&+&+&+&0\end {array} \right].
\nonumber
\end{eqnarray}
Here, the matrix $\gamma^{(m)}$ denotes the matrix $\gamma^{(\sigma)}$
for the permutation No.~$m$ from the Table \ref{table:numerics}.

The results of Monte Carlo numeric evaluation of the seven-fold
integrals $K^{(\sigma)}$ given by Eq.~(\ref{K-sigma}) are summarized
in Table \ref{table:numerics}. Performing summation with the
multiplicities $M^{(\sigma)}$ we get finally the estimate for the
coefficient $d_4$ in the Taylor expansion (\ref{D-series}):
\be
  d_4 = - (7 \pm 7) \times 10^{-5}.
\label{d4-result}
\ee
The numerical uncertainty indicated in Eq.~(\ref{d4-result})
corresponds to one standard deviation.
Thus, we cannot distinguish $d_4$ from zero and can estimate
the upper bound for its absolute value as $|d_4|<3\times10^{-4}$.

\section{Arbitrary dependence of $\vp(t)$}
\label{section-phi(t)}

In this section we discuss to what extent the results obtained above
can be generalized to an arbitrary time dependence of the control
parameter $\vp(t)$.

We start by summarizing the modifications of the theory introduced
by an arbitrary $\vp(t)$. For a generic $\vp(t)$,
the diffuson becomes a function of {\em three}\/ times
[cf.\ Eq.~(\ref{full-diffuson-definition})]:
\be
  \langle  Q^{(+)}_{t+\eta/2,t-\eta/2} Q^{(-)}_{t'-\eta'/2,t'+\eta'/2} \rangle
  = \frac{2\Delta}{\pi}
  \delta(\eta-\eta')
  {\cal D}_{\eta}(t,t') .
\label{full-diffuson-phi}
\ee
To study the action (\ref{action-general}) one can develop the standard
perturbation theory described in Section \ref{section-diagrams}. In terms
of the $b$-fields, the action will have the form (\ref{S[b]}) with the
quadratic part ($\vp_i\equiv\vp(t_i)$)
\be
S^{(2)}[b,\bar{b}] = \frac{\pi}{2\Delta} \int\!\!\int dt_1\, dt_2\,
\bar{b}_{12}\Big[ (\partial_1 + \partial_2) + \Gamma(\vp_1-\vp_2)^2 \Big]
b_{21}
\label{S2[b]-general}
\ee
and infinite number of nonlinear terms $S^{\geq4}[b,\bar b]$.
In Eq.~(\ref{S2[b]-general}) we introduced $\Gamma=\pi\Delta C(0)$.
The bare diffuson defined as the propagator of $S^{(2)}[b,\bar b]$
is given by~\cite{AAK1982,VavilovAleiner1999,SBK04}
\be
  {\cal D}^{(0)}_{\eta}(t,t')
  = \theta(t-t')
  \exp\left\{
  - \Gamma \int_{t'}^t [\vp(\tau+\eta/2)-\vp(\tau-\eta/2)]^2 d\tau
  \right\} .
\label{bare-diffuson-phi}
\ee

It is remarkable that with such a modification of the theory one can still
prove the cancellation of all {\em internal}\/
vertices of the order higher than 4
in the rational parameterization.
Indeed, the proof presented in Section \ref{section-cancellation}
was based on the equation of motion (\ref{eq-motion}) for the diffuson,
which is now replaced by an analogous equation
\be
  \partial_t {\cal D}^{(0)}_\eta (t,t')
  = \delta(t-t') - \Gamma [\vp(t+\eta/2)-\vp(t-\eta/2)]^2
  {\cal D}^{(0)}_\eta (t,t') ,
\label{eq-motion-general}
\ee
and the combinatorial counting of coefficients which is insensitive
to time dependence of $\vp(t)$.

The resulting $b$-theory has the action
\be
  S = S^{(2)}[b,\bar{b}]
  -
  \frac{\pi\Gamma}{8\Delta} \int dt_1 dt_2 dt_3 dt_4\,
  (\vp_1-\vp_2)(\vp_3-\vp_4) \,
  b_{12}\bar{b}_{23}b_{34}\bar{b}_{41} ,
\label{b-theory-general}
\ee
which is a generalization of Eq.~(\ref{b-action-rational})
for an arbitrary dependence of the control parameter $\vp(t)$.
Thus, the diagrams for the pro-diffuson $\corr{b\bar{b}}$:
\be
  \langle b_{t+\eta/2,t-\eta/2} \bar{b}_{t'-\eta'/2,t'+\eta'/2} \rangle
  = \frac{2\Delta}{\pi}
  \delta(\eta-\eta')
  {\cal D}^*_{\eta}(t,t')
\label{star-diffuson-phi}
\ee
with an arbitrary $\vp(t)$ are exactly the same as the diagrams
in the linear case $\vp(t)=vt$ considered in the previous Sections
(but with the new diffusons (\ref{bare-diffuson-phi})).

The energy absorption rate is expressed through the full diffuson
${\cal D}_\eta(t,t')$ defined in terms of the field $Q$
by Eq.~(\ref{full-diffuson-phi}).
It differs from the pro-diffuson ${\cal D}^*_\eta(t,t')$ in the $b$-theory
since $Q$ is a nonlinear function of $b$ and $\bar b$.
In studying the linear perturbation this difference was irrelevant
for the calculation of the diffusion coefficient (see Section
\ref{section-diagrams-D}). For a generic perturbation $\vp(t)$,
the difference between the full diffuson ${\cal D}_\eta(t,t')$
and the pro-diffuson ${\cal D}^*_\eta(t,t')$ becomes important.
In particular, the two-loop analysis of the quantum interference
correction under the action of a harmonic perturbation~\cite{SBK04}
has shown that the average $\corr{b\bar{b}b\cdot \bar{b}}$
(the diagram (c) in Ref.~\onlinecite{SBK04}) has a contribution
comparable to that of the average $\corr{b\cdot \bar{b}}$,
both being negative and growing with time $\propto t$.

\section{Relation to the Dyson-Maleev transformation}
\label{section-DM}

In this section we elucidate the meaning of the effective
$b$-theory (\ref{b-action-rational})
and discuss its relation to the Dyson-Maleev
transformation widely used in dealing with quantum spin ferromagnets.

The action (\ref{b-action-rational}) has been obtained in the previous sections
from an analysis of mutual cancellations of higher-order vertices
in the perturbation theory for the rational parameterization.
However, it turns out that the same action may be obtained directly
from the initial $\sigma$-model action (\ref{action-sigma-model})
if one adopts the following
parameterization of the $Q$-matrix in terms of the fields $b$ and $\bar b$:
\be
  Q =
  \begin{pmatrix}
    1-b\bar b/2 && b-b\bar bb/4 \\
    \bar b && -1+\bar bb/2
  \end{pmatrix}.
\label{Q-DM}
\ee
Indeed, this parameterization respects the nonlinear constraint $Q^2=1$, and
the trivial algebra of substituting (\ref{Q-DM}) into the initial
$\sigma$-model action (\ref{action-sigma-model})
immediately leads to the effective action (\ref{b-action-rational})
of the $b$-theory. With the explicit parameterization (\ref{Q-DM}),
{\em external}\/ vertices (matrix elements of the $Q$-matrix)
also become finite polynomials in $b$ and $\bar b$.
We have checked that the two approaches (direct use of the
parameterization (\ref{Q-DM}) and the vertex cancellation
by the technique developed in Appendix \ref{A:rational})
produce the same effective external vertices.

Thus we come to an important conclusion about the $\sigma$-model
(\ref{action-sigma-model}):
The rational parameterization (\ref{parameter-rational})
(which contains an infinite series
of higher-order interaction vertices) after
mutual cancellation of higher-order vertices
in the diagrammatic expansion is perturbatively
equivalent to the parameterization (\ref{Q-DM}).

In the rational parameterization [more generally, in any
parameterization of the form (\ref{W-b})--(\ref{parameter-general-2})],
the matrices $b$ and $\bar b$
are Hermitian conjugate of each other: $\bar b=b^\dagger$.
The same remains therefore true for the $b$-theory (\ref{b-action-rational})
obtained from the rational parameterization.
On the other hand, the parameterization (\ref{Q-DM})
with $\bar b=b^\dagger$
violates the hermiticity of the $Q$ matrix.
Nevertheless, our derivation indicates
that this violation is inessential at the perturbative level
and can be taken into account by a proper deformation of the integration
contour over the elements of the $Q$ matrix.

The parameterization (\ref{Q-DM}) is closely related to
the famous Dyson-Maleev \cite{Dyson56,Maleev57} parameterization
for quantum spins. In that representation, the spin-$S$ operators
are expressed by the boson creation and annihilation
operators $\hat a^\dagger$ and $\hat a$ as
\be
  \hat S^+ = (2S-\hat a^\dagger \hat a) \hat a,
\qquad
  \hat S^- = \hat a^\dagger,
\qquad
  \hat S^z = S-\hat a^\dagger \hat a .
\label{DM}
\ee
The Dyson-Maleev transformation conserves the spin
commutation relations but violates the property
$(\hat S^-)^\dagger=\hat S^+$
rendering the spin Hamiltonian manifestly non-Hermitian.
This is the expense one has to pay for making
the spin Hamiltonian a finite-order polynomial
in boson operators [as opposed to an infinite
series in the Holstein-Primakoff
parameterization \cite{HolstPrim40};
note that the Holstein-Primakoff parameterization is analogous
to the square-root-even parameterization (\ref{param-sqrt-even})].
The Dyson-Maleev transformation has proven to be the most convenient tool
for studying spin wave interaction \cite{Canali92,Hamer92-93}:
it reproduces all the
perturbative results obtained with the Holstein-Primakoff
parameterization in a much faster and compact way.

A classical analogue of the Dyson-Maleev transformation
was recently used by Kolokolov \cite{Kolokolov2000}
in studying two-dimensional classical ferromagnets.
To the best of our knowledge,
there exists just one article by
Gruzberg, Read and Sachdev \cite{Gruzberg1997}
where the Dyson-Maleev parameterization
was applied for the perturbative treatment of a $\sigma$-model
(in the replica form).

In the Dyson-Maleev representation,
the Hilbert space of free bosons should be truncated
in order for the operator $\hat S^z$ to have a bounded
spectrum. In Ref.~\onlinecite{Kolokolov2000}, this truncation
corresponds to integrating over a bounded region in bosonic
variables. We expect that a similar constraint on the
matrices $b$ and $\bar b$ may be required in order to achieve
a nonperturbative equivalence between the initial
$\sigma$-model in the $Q$-representation and the $b$-theory.
This question is of importance for studying non-perturbative
effects in $\sigma$-models, but goes beyond the scope of the present paper.

\section{Discussion}
\label{section-conclusion}

This work has appeared as a result of our attempt to prove the conjecture
formulated in Ref.~\onlinecite{skvor} that the Kubo formula for the
energy absorption rate of a linearly driven unitary random
Hamiltonian gives an exact result in the whole range of driving velocities.
In the notation of the present paper, this conjecture implies that
the relation $D(\alpha)=\alpha$ is exact.

Being unable to verify the conjecture nonperturbatively,
we calculate the four-loop correction to the Kubo
formula $D(\alpha)=\alpha$ in the limit of large velocities
of the driving field, $\alpha\gg 1$.
In the process of our derivation,
we have refined the $\sigma$-model approach of Ref.~\onlinecite{skvor}:
our improved method does not involve the fermionic distribution function
and expresses the energy diffusion coefficient (\ref{diffusion-coefficient})
in terms of the full diffuson (\ref{full-diffuson-definition})
of the field theory (\ref{action-sigma-model}).

We have further proven that the resulting $\sigma$-model is
{\em perturbatively equivalent}\/ to the specific matrix $\phi^4$-theory
(\ref{b-action-rational}).
The lack of higher nonlinearities makes it
possible to classify all four-loop diagrams and write down analytic
expressions for them without resorting to computer symbolic computations.
The final evaluation of emerging 7-dimensional integrals (\ref{K-sigma})
cannot be done analytically, and we calculate them numerically.
We find that the coefficient $d_4$ in the expansion (\ref{D-series})
is indistinguishable from zero within the precision of our calculation,
with its absolute value bounded by $|d_4|<3\times10^{-4}$.

We believe that this conclusion acts in favor of the conjecture
$D(\alpha)=\alpha$.

This result may appear less surprising
if we recall that the static (time-independent)
unitary random-matrix ensemble is also known to possess
some peculiar properties.
In particular, its spectral statistics can be mapped onto
the problem of one-dimensional noninteracting fermions;
for a certain class of integrals over the unitary group
the saddle-point approximation is exact
(the Duistermaat-Heckman theorem~\cite{Duistermaat-Heckman},
see Ref.~\onlinecite{Zirnbauer99} for a discussion).
However, in the time-dependent problem considered in
the present work, we are not able to perform mapping
onto free fermions,
and the Duistermaat-Heckman theorem does not help to evaluate the
functional integral of the Keldysh $\sigma$-model.
The $\sigma$-model is nontrivial, and its diffuson
${\cal D}_\eta(t)$ is a complicated function of $\eta$ and $t$
at arbitrary value of the coupling $\alpha$.
The diffuson ${\cal D}_\eta(t)$ is
renormalized from the simple diffusive form (\ref{bare-diffuson}),
and our result indicates that only its long-time
asymptotics [which determines the energy diffusion
coefficient (\ref{diffusion-coefficient})] is free from perturbative
corrections up to the order $\alpha^{-4/3}$.

The verification of the original nonperturbative conjecture $D(\alpha)=\alpha$
remains an open question.

As a byproduct of our analysis, we have rederived the $Q$-matrix
parameterization (\ref{Q-DM})
which is analogous to the Dyson--Maleev parameterization
for spin operators \cite{Gruzberg1997}.
The parameterization (\ref{Q-DM}) is not specific to the Keldysh formalism,
and can be applied to a wide class of {\em unitary}\/ $\sigma$-models
(e.g., to that describing diffusion of a particle in random media,
both in the supersymmetric or replica \cite{Gruzberg1997} approaches),
leading to a considerable simplification of the perturbative expansion.
We could not extend this parameterization to the orthogonal and symplectic
$\sigma$-models, since those involve additional linear constraints on the
$Q$-matrix, apparently incompatible with our parameterization.

\acknowledgments

We thank M. V. Feigel'man and I. V. Kolokolov for drawing
our attention to the Dyson-Maleev parameterization,
and I. Gruzberg for pointing out Ref.~\onlinecite{Gruzberg1997}.
This research was partially (M.~A.~S.)
supported by the Program ``Quantum Macrophysics''
of the Russian Academy of Sciences,
RFBR under grant No.\ 04-02-16998,
the Dynasty foundation and the ICFPM.
M.~A.~S. acknowledges the hospitality of the
Institute for Theoretical Physics at EPFL,
where the main part of this work was performed.

\appendix

\section{Ward identity and energy absorption rate}
\label{A:Keldysh}

In this Appendix we show that in the many-particle formulation
the energy absorption rate under the action of an arbitrary perturbation
$\vp(t)$ is expressed through the generalized diffuson
(\ref{full-diffuson-phi}). In particular, for the linear perturbation
$\vp(t)=vt$, the diffusion coefficient in the
energy space may be extracted from the decay rate of the diffuson
(\ref{full-diffuson-definition}). As a byproduct of our discussion,
we also prove that, at $\eta=0$, the diffuson reduces to the step function:
\be
  {\cal D}_{\eta=0}(t)=\theta(t)\, .
\ee

\subsection{Two approaches to energy diffusion}

The diffusion coefficient in the energy space may be defined in two
different ways.

In the original sigma-model derivation~\cite{skvor}, the
states of the time-dependent Hamiltonian $H(t)$ were occupied by
noninteracting fermions. In such a multi-particle formulation,
the step-like structure of the fermionic distribution function $f(E)$,
\be
  \lim_{E\to-\infty}f(E)=1, \qquad \lim_{E\to\infty}f(E)=0,
\label{fermi-bc}
\ee
generates a spectral flow of fermions from low to high energies
leading to the increase of the total energy of the system with time.
At large time and energy scales, this many-particle process
may be described by a diffusion equation on the distribution
function $f(E)$. The corresponding diffusion coefficient expressed
as $D\Delta^3$ (where $\Delta$ is the average interlevel spacing and
$D$ is dimensionless) translates into the energy pumping rate
$W=D\Delta^2$ (the density of states is $\Delta^{-1}$).

On the other hand, the same diffusion process may be observed in a
single-particle problem. In the single-particle quantum mechanics
(\ref{schroedinger}), the energy of the system $E(t)$ defined with
the instantaneous Hamiltonian
\be
 E(t)=\langle \Psi(t)|H(t)|\Psi(t)\rangle
\ee
diffuses with time as described by (\ref{diffusion-definition}).
One can expect that in the process of time evolution of $H(t)$, the relative
phases of the wave function components corresponding to widely
separated energies become uncorrelated, and therefore the multi-particle
diffusion evolution of $f(E)$ and the single-particle diffusion process
give equivalent definitions of the diffusion coefficient $D$.
The above reasoning is justified for a non-periodic evolution of $H(t)$
(for example, for the linear evolution of the parameter $\varphi(t)=vt$);
for a periodic perturbation studied in Ref.~\onlinecite{BSK03}, the phase
correlations become important which leads to dynamic localization. Nevertheless,
we show below that the energy pumping rate $W$ in the multi-particle
problem may be expressed in terms of the single-particle diffuson
for a rather general time dependence $\varphi(t)$.

\subsection{Ward identity}

A helpful tool for our further derivation are the Ward identities
generated by rotations of the integration variables $Q$ in the
functional integral of the sigma-model.

Consider the  functional
\be
  \Pi[V] = \int V^{-1}QV \, e^{-S[V^{-1}QV]} [DQ] ,
\label{Ward0}
\ee
where $V$ is an arbitrary matrix in the time and Keldysh spaces
(not necessarily unitary), and the action is given
by Eq.~(\ref{action-general}). The integration is
performed over the matrices $Q$ of the form (\ref{orbit}), (\ref{lambda-def}).
If the rotation by the matrix
$V$ is local (with $V_{tt'}\to\delta_{tt'}$ at $|t|,|t'|\to\infty$), it
may be compensated by changing the
integration variable $Q\mapsto V^{-1}QV$ in Eq.~(\ref{Ward0}) producing
no anomalous contribution, and therefore $\Pi[V]$ is independent of
such rotations $V$.

The multi-particle approach described in the previous subsection
may also be introduced with the same rotated path integral (\ref{Ward0}),
but with the {\it anomalous}\/ rotation matrix $V$ (involving the distribution
function $f(E)$)~\cite{skvor}. This anomalous rotation will be discussed
in detail below in the subsequent subsection, and in this subsection we
derive the Ward identity generated by infinitesimal non-anomalous rotations
$V$.

Expanding $V=1+A+\dots$ and taking the variation of $\Pi[V]$
with respect to the infinitesimal generator $A$ we get
\be
  \delta\Pi =
  [\corr{Q},A]
  + \frac{\pi i}{\Delta} \corr{Q \Tr ([Q,\hat E] A)}
  + \pi^2 C(0) \corr{Q \Tr ([\hat\vp, Q \hat\vp Q] A)
 }
  = 0 \, .
\label{Ward1}
\ee
In deriving the second term with $[Q,\hat E]$ we performed a cyclic
permutation under the trace, which is equivalent to integration by parts
and omitting the resulting boundary terms. This procedure is justified
for local $A_{tt'}$.

Various components of Eq.~(\ref{Ward1}) associated with different elements
of the matrix $A$ give the set of Ward identities related to
the $V$-invariance of $\Pi[V]$.
Of particular importance is its off-diagonal (Keldysh) component $\Pi^{(+)}$
generated by the Keldysh element $A^{(+)}$:
\be
  \delta_{t_1t_4}\delta_{t_2t_3}
  + \delta_{t_1-t_2,t_4-t_3}
    (\partial_3+\partial_4)
      \, {\cal D}_{t_1-t_2} \! \left( \frac{t_1+t_2}{2},\frac{t_3+t_4}{2} \right)
  + \frac{\pi^2}{2} C(0) (\vp_3-\vp_4) \int dt_5 \, \vp_5 \,
    \corr{Q^{(+)}_{12} (Q_{35}Q_{54})^{(-)}
  } = 0 ,
\label{Ward2}
\ee
where the diffuson with three times ${\cal D}_\eta(t,t')$
is defined in Eq.~(\ref{full-diffuson-phi}) for an arbitrary
time dependence $\varphi(t)$, and we have used that,
due to causality, $\corr{Q_{tt'}}=\Lambda=\sigma_3\delta_{tt'}$.

\subsection {Causality of the diffuson}

The diffuson ${\cal D}_\eta(t,t')$ may be calculated by summing
the diagrammatic series as described in Section \ref{section-diagrams}.
Every line in the diagram is the bare retarded diffuson
(\ref{bare-diffuson-phi}), and therefore the full diffuson
${\cal D}_\eta(t,t')$ also equals zero when $t<t'$.

Furthermore,
using the Ward identity (\ref{Ward2}) at $\eta=0$, we immediately find that
the diffuson ${\cal D}_\eta(t,t')$ reduces to the step function of $t-t'$.
Indeed, setting $t_3=t_4$ nullifies the last term yielding
$\partial_{t'} {\cal D}_{\eta=0}(t,t') = -\delta(t-t')$.
Integrating this equation and using the causality of the diffuson
(${\cal D}_\eta(t,t')=0$ for $t<t'$), we obtain
${\cal D}_{\eta=0}(t,t')=\theta(t-t')$.

For our discussion in the next subsection, in order to regularize the
diffusion process at $t \to -\infty$, we consider the situation
where the evolution of the Hamiltonian $H(t)$ switches on at a certain
time moment $t_0$. This can be modelled by a time dependence
of $\vp(t)$ such that it remains constant at earlier times
$\vp(t<t_0)={\rm const}$. For constant $\vp(t)$, the last term in
(\ref{Ward2}) vanishes and (similarly to the case $\eta=0$) we obtain
\be
\frac{\partial}{\partial t'}
{\cal D}_\eta(t,t') + \delta(t-t') =0
\qquad
{\rm for} \quad t'<t_0-\frac{|\eta|}{2}.
\ee
We can integrate this equation in two domains:
\be
{\cal D}_\eta(t,t') = \theta(t-t')
\qquad
{\rm for} \quad t,t'<t_0-\frac{|\eta|}{2} ,
\label{caus-1}
\ee
and
\be
{\cal D}_\eta(t,t') = {\cal D}_\eta(t,-\infty)
\qquad
{\rm for} \quad t'<t_0-\frac{|\eta|}{2}<t\, .
\label{caus-2}
\ee
The first property (\ref{caus-1}) guarantees that the diffuson
``switches on'' only after the moment $t_0$ when the evolution of the
$H(t)$ starts. The second property (\ref{caus-2}) states that the
diffuson with the initial time before $t_0$ does not actually
depends on this initial time, which allows us to define the
diffuson originating at $t=-\infty$.

\subsection{Distribution function and energy absorption rate}

To prove the relation between the multi-particle and single-particle
definitions of the energy diffusion (see the first subsection of
this Appendix), we consider the multi-particle kinetic problem
in which the system is initially in a stationary state
characterized by an arbitrary fermionic distribution function $f^{(0)}(E)$.
The evolution of the Hamiltonian starts at a time moment $t_0$, which
is described by $\vp(t<t_0)={\rm const}$.
Within the Keldysh formalism~\cite{RammerSmith1986,KamenevAndreev99},
occupation of states by non-interacting
fermions may be taken into account by rotating the matrix
$Q$ by the upper-triangular matrix~\cite{UF-comment}
\be
  (V_{F^{(0)}})_{tt'} =
  \begin{pmatrix}
      \delta_{tt'} & F_{tt'}^{(0)} \\
      0 & \delta_{tt'}
  \end{pmatrix} ,
\label{UF}
\ee
where $F_{tt'}^{(0)}$ is the Fourier transform (in $t-t'$) of
$F^{(0)}(E)=1-2f^{(0)}(E)$.
The evolution
of the distribution function with time may be read off as
\be
  F=\frac{1}{2} \Pi^{(+)}[V_{F^{(0)}}] .
\ee
Note that $V_{F^{(0)}}$ is not a local rotation (it does not vanish
at $t\to -\infty$), and therefore $F$ undergoes a non-trivial time
evolution.

We may calculate the time evolution of $F$ in the same procedure as
the derivation of the Ward identity (\ref{Ward2}).
The only difference is that the second term in Eq.~(\ref{Ward1})
originating from $S_\text{E}[V^{-1}QV]$ vanishes for $V=V_{F^{(0)}}$.
This can be most easily seen in the energy representation,
where $F^{(0)}$ is diagonal and evidently commutes with $\hat E$.
Then, applying the Ward identity (\ref{Ward2}) one arrives at
the result
\be
  \Pi^{(+)}[V_{F^{(0)}}]_{12}
  =
  - 2 \int dt_3\, dt_4\,
  F^{(0)}_{t_4-t_3} \,
  \delta_{t_1-t_2,t_4-t_3}
    (\partial_3+\partial_4)
      \, {\cal D}_{t_1-t_2} \!
      \left( \frac{t_1+t_2}{2},\frac{t_3+t_4}{2} \right) ,
\ee
which translates into
\be
  F_{t+\eta/2,t-\eta/2}
  = {\cal D}_{\eta}(t,-\infty) F^{(0)}(\eta) .
\label{F2}
\ee
This equation proves that the diffuson ${\cal D}_{\eta}(t,t')$ formally
defined as a $\corr{QQ}$ correlation function in the field
theory (\ref{action-general}) indeed plays the role of the
evolution kernel for the distribution function $F(t,E)$.

The energy absorption rate can be expressed in terms of
the distribution function as~\cite{skvor,BSK03}
\be
  W(t) =
  - \frac{\pi i}{\Delta} \lim_{\eta\to0}
    \partial_{t}\partial_{\eta} F_{t+\eta/2,t-\eta/2} .
\ee
Using Eq.~(\ref{F2})
and the asymptotics $F^{(0)}(\eta)\sim1/(i\pi\eta)$ at $\eta\to0$
which follows from the fermionic boundary conditions (\ref{fermi-bc}),
we get
\be
  W(t) = - \frac1{2\Delta} \left. \frac{\partial}{\partial t}
    \frac{\partial^2}{\partial \eta^2} \right|_{\eta=0}
    {\cal D}_\eta(t,-\infty) .
\label{W-general}
\ee

If we consider the situation of a linear time evolution $\vp=vt$
switched on at a certain time moment $t_0$, the linearly growing
contribution to the three-time diffuson  ${\cal D}_\eta(t,-\infty)$
equals that of the two-time diffuson ${\cal D}_\eta(t-t_0)$ of
the translationally invariant theory (\ref{full-diffuson-definition}).
This can be easily seen from the diagrammatic expansion of the
diffuson ${\cal D}_\eta(t,-\infty)$: the characteristic decay time
of diagrams is $t^*=\Delta^{-1}\alpha^{-1/3}$, and switching
on at time $t_0$ plays the role of a soft cut-off for the vertices
(\ref{b-theory-general}). Therefore, at time scales $t-t_0\gg t^*$,
the details of the switching-on process become unimportant, and
we may replace  ${\cal D}_\eta(t,-\infty)$ in (\ref{W-general})
by ${\cal D}_\eta(t-t_0)$. This proves the equivalence of our
single-particle definition (\ref{diffusion-coefficient})
of the diffusion coefficient
to the earlier multi-particle approach of Refs.~\onlinecite{BSK03,skvor,SBK04}.

\section{Parameterizations of the $Q$ matrix}
\label{A:parameterizations}

\subsection{The Jacobian}

In this Appendix we discuss different parameterizations of
matrices $Q$ by elements $W$ of the corresponding tangent space, according to
(\ref{parameter-general-2}). With such parameterizations, the integration over the
non-linear manifold of matrices $Q$ reduces to that over the linear space of $W$
(i.e., over the fields $b_{tt'}$ and ${\bar b}_{tt'}$):
\be
  \int[DQ] = \int e^{-S_J(W)} [DW] ,
\ee
where $J_f[W]=\exp[-S_J(W)]$ is the Jacobian depending
on the function $f(W)$ in (\ref{parameter-general-2}).

The integration measure $[DQ]$ may be defined from the invariant metric on the
space of $Q$ matrices:
\be
(dQ,dQ) \equiv -\Tr dQ\, dQ .
\ee
Similarly, the integration measure $[DW]=[Db\, D{\bar b}]$ may be defined with the
metric
\be
(dW,dW) \equiv -\Tr\, dW\, dW = 2 \Tr\, db\, d{\bar b} .
\ee
The Jacobian $\exp[-S_J(W)]$ is the square root of the determinant of the metric $(dQ,dQ)$ relative
to $(dW,dW)$. Further in the Appendix, we compute explicitly the determinant action
$S_J(W)$ for any given parameterization $f(W)$ and find those parameterizations
which have the unit Jacobian ($S_J(W)=0$).

\bigskip

The function $f(W)$ satisfies the nonlinear constraint:
\be
  f(W) f(-W)=1 .
\label{W-constraint}
\ee
This constraint is solved by the condition that $\ln f(W)$
is an odd function of $W$.

The function $f(W)$ must be regular at $W=0$, with the Taylor
expansion
\be
f(W)=1+c_1 W + c_2 W^2 + \dots
\label{appendix-f-expansion}
\ee
Without loss of generality, we can normalize $W$ by the condition
$c_1=1$. Then necessarily $c_2=1/2$.
Other coefficients have a certain freedom.
For example, one can chose $c_{2n+1}$ arbitrarily,
then $c_{2n+2}$ are uniquely determined by the
constraint (\ref{W-constraint}).
We naturally define $c_0=1$.

The quadratic form $(dQ,dQ)$ may now, using the cyclic trace property,
be written as
\be
(dQ,dQ)= - \Tr\, df(W)\, df(-W) =
\sum_{n_1,n_2} \alpha_{n_1,n_2} \Tr(W^{n_1}\, dW\, W^{n_2}\, dW)
\ee
with symmetric coefficients $\alpha_{n_1,n_2}=\alpha_{n_2,n_1}$.
Using the tensor-product notation, we can write this as
\be
(dQ,dQ)=(dW\otimes dW, {\bf R}),
\label{metric-tensor-product}
\ee
where
\be
{\bf R}= \sum_{n_1,n_2} \alpha_{n_1,n_2} W^{n_1} \otimes W^{n_2}
\ee
and $( \cdot , \cdot )$ in the right-hand side of (\ref{metric-tensor-product})
is understood as $(A \otimes B, C \otimes D) = \Tr (ACBD)$.
The Jacobian equals $(\det {\bf R})^{1/2}$ and can be generated as the
``Jacobian action'' $S_J(W)$:
\be
  S_J = -\frac{1}{2} \Tr \ln {\bf R} .
\ee

The final step of the derivation is to express ${\bf R}$ in terms of $f$.
Since ${\bf R}$ contains only commuting matrices $W$, it
is convenient to replace ${\bf R}$ by the function $R(x,y)$
defined as
\be
R(x,y) = \sum_{n_1,n_2} \alpha_{n_1,n_2} x^{n_1} y^{n_2} ,
\ee
so that
\be
  {\bf R} = R(W \otimes {\bf 1},{\bf 1} \otimes W) .
\ee
By inspection,
\be
\alpha_{n_1,n_2}=\sum_{k_1=0}^{n_1} \sum_{k_2=0}^{n_2}
(-1)^{k_1+k_2} c_{k_1+k_2+1} c_{n_1+n_2-k_1-k_2+1} .
\ee
In terms of generating functions $f(x)$ and $R(x,y)$,
this can be written as
\be
R(x,y)= - \frac{f(x)-f(y)}{x-y}\, \frac{f(-x)-f(-y)}{x-y}
 = \left[\frac{\sqrt{\frac{f(x)}{f(y)}}-\sqrt{\frac{f(y)}{f(x)}}}{x-y}\right]^2 .
\ee

Now we can write down the explicit formula for $S_J(W)$ in terms of $f(W)$.
For a given $f(W)$, we may expand
\be
\ln \frac{\sqrt{\frac{f(x)}{f(y)}}-\sqrt{\frac{f(y)}{f(x)}}}{x-y} =
\sum_{n_1,n_2} \beta_{n_1,n_2} x^{n_1} y^{n_2} ,
\label{beta-definition}
\ee
and then
\be
  S_J(W)= - \sum_{n_1,n_2} \beta_{n_1,n_2} \Tr(W^{n_1}) \Tr(W^{n_2}) .
\label{Jacobian-action}
\ee

\subsection{Potentially useful parameterizations}

For doing perturbative calculation in the sigma-model
one has to chose a certain parameterization.
We want to mention four possibilities:
\begin{itemize}
\item
{\em Square-root-even}\/ parameterization
(in this parameterization, $f(W)$ does not contain
odd powers of $W$ higher than one):
\be
  f(W) = \sqrt{1+W^2}+W .
\label{param-sqrt-even}
\ee
This parameterization is widely used in literature since
perturbative expansion in this parameterization directly
corresponds~\cite{HikamiBox} to the standard diagrammatic technique
with cross averaging over disorder~\cite{diagrammatics}.
The Jacobian in the square-root-even parameterization in not equal to 1.

\item
{\em Exponential}\/ parameterization:
\be
  f(W) = \exp(W) .
\ee
This form was used by Efetov
to construct his famous parameterization~\cite{Efetov1983}
of the integration manifold for the zero-dimensional
supersymmetric sigma-model, that opened a way to exact evaluation
of the two-level correlation function.
The Jacobian in the exponential parameterization is not equal to 1.

\item
{\em Rational}\/ parameterization:
\be
  f(W) = \frac{1+W/2}{1-W/2} .
\label{rational}
\ee
This parameterization was suggested by Efetov~\cite{Efetov1983}
in studying the supersymmetric sigma-model.
The rational parameterization is frequently used
for its Jacobian is equal to 1~\cite{Efetov1983,Efetov-book}.

\item
{\em Square-root-odd}\/ parameterization
(in this parameterization, $f(W)$ does not contain
even powers of $W$ higher than two):
\be
  f(W)
    = 1+\frac{W^2}{2} + W \sqrt{1+\frac{W^2}{4}}
    = \left(\sqrt{1+\frac{W^2}{4}}+\frac{W}{2} \right)^2 .
\label{param-sqrt-odd}
\ee
This parameterization was used in Refs.~\onlinecite{VerbZirn85,YurLerner99,Zirnbauer99}
for the study of the level statistics of random Hamiltonians within
the replica formalism. The square-root-odd parameterization
is known to have Jacobian equal one.
In addition to that, the choice of the parameterization (\ref{param-sqrt-odd})
renders the action of the zero-dimensional sigma-model
$S[Q]\propto\Tr\Lambda Q$ to be Gaussian in $W$.
\end{itemize}

Below we shall prove that the class of parameterizations
with unit Jacobian (e.g., $S_J(W)=0$)
consists of a one-parametric family (\ref{unit-Jacobian-parameter}),
with the {\em rational}\/ and {\em square-root-odd}\/
parameterization being particular cases.

\subsection{Parameterizations with unit Jacobian}

Now we shall solve the problem of finding all parameterizations $f(W)$ leading to the
trivial Jacobian $S_J(W)=0$. Some powers of $W$ in the Jacobian action (\ref{Jacobian-action})
give zero traces: $\Tr(W^0)=0$ due to the causality and $\Tr(W^n)=0$ for all odd $n$.
Therefore we look for functions $f(x)$ producing in (\ref{beta-definition})
only nonzero $\beta_{n_1,n_2}$ with one of the indices zero or odd.

Thus, a parameterization with unit Jacobian must satisfy the condition
\be
\ln \frac{\sqrt{\frac{f(x)}{f(y)}}-\sqrt{\frac{f(y)}{f(x)}}}{x-y} =
F_1(x) + F_1(y) + F_2(x,y)
\label{unit-J-cond}
\ee
with antisymmetric $F_2(x,y)=F_2(-x,-y)=-F_2(-x,y)=-F_2(x,-y)$.
By setting $x=0$ or $y=0$ we obtain
\be
  F_1(x) = \ln \frac{f^{1/2}(x)-f^{-1/2}(x)}{x} .
\ee
Finally, by symmetrizing the left-hand side of Eq.~(\ref{unit-J-cond}),
we get rid of $F_2(x,y)$ and obtain a closed equation on $f(x)$:
\be
\frac{[f(x)+f(-x)] - [f(y)+f(-y)]}{x^2-y^2} =
\frac{[f(x)+f(-x)-2][f(y)+f(-y)-2]}{x^2 y^2} .
\ee
The latter equation may be simply solved as
\be
f(x)+f(-x) = 2 + \frac{x^2}{1-\lambda\frac{x^2}{4}}
\ee
with an arbitrary constant $\lambda$. For $f(x)$ this gives
\be
  f(x) =
  \frac{\left(\frac{x}{2} + \sqrt{1+(1-\lambda)\frac{x^2}{4}}\right)^2}
  {1-\lambda\frac{x^2}{4}} .
\label{unit-Jacobian-parameter}
\ee
Thus the parameterization with the unit Jacobian form a one-parameter family.
Out of the four examples of parameterizations mentioned above, two belong to this
family: at $\lambda=1$, the expression (\ref{unit-Jacobian-parameter})
gives the {\it rational} parameterization, and at
$\lambda=0$ it gives the {\it square-root-odd} parameterization.
The {\it square-root-even} and {\it exponential} parameterizations
do not belong to this family and have
non-trivial Jacobian contributions $S_J(W)$.

The main calculation of the paper (cancellation of higher-order vertices) is done
in the rational parameterization. We could not obtain similar results in other
parameterizations.

\section{Cancelation of higher-order vertices in rational parameterization}
\label{A:rational}

\subsection{Combinatoric coefficients and diagrammatic rules}

The goal of this section is to show that all internal vertices
of order higher than four cancel in the rational parameterization.

\begin{figure}
\epsfxsize=0.5\hsize
\centerline{\epsfbox{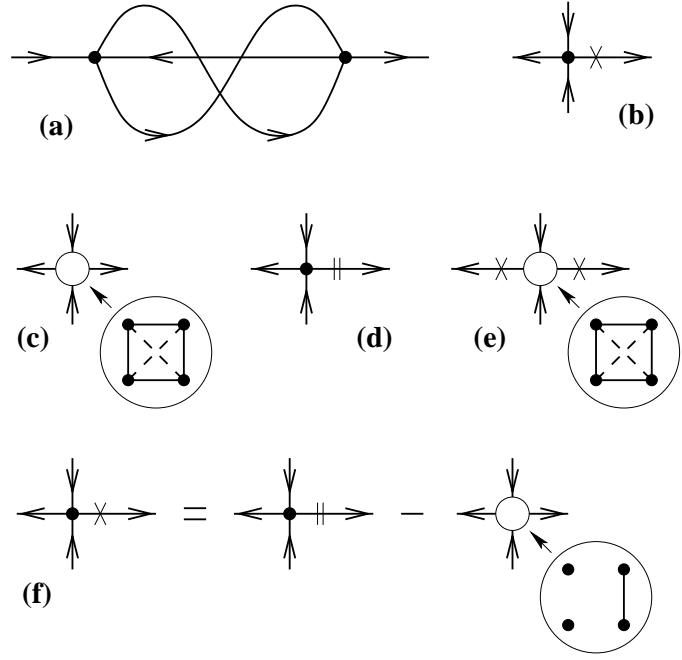}}
\medskip
\caption{
{\bf (a)} The same diagram as in Fig. \protect\ref{fig:diagram-examples}c
in the ``thin-line'' notation.
{\bf (b)} Differentiation $\partial_t {\cal D}^{(0)}_\eta(t)$.
{\bf (c)} Vertex with $(\varphi_i-\varphi_j)^2$ terms.
At the vertex, the solid/dashed lines denote $a_{ij}=\pm 1$ respectively,
cf.\ Eq.~(\ref{a-ij}).
{\bf (d)} Contraction of the diffuson: ${\cal D}^{(0)}_\eta(t)$ is replaced by $\delta(t)$.
{\bf (e)} An original vertex of the sigma-model action (four-valent vertex
is shown as an example). The differentiations $\hat\partial$ and
$a_{ij}(\varphi_i-\varphi_j)^2$ terms are {\it added}\/ together.
{\bf (f)} Graphical presentation of the diffuson equation
of motion (\protect\ref{eq-motion}):
$\partial_t {\cal D}^{(0)}_\eta(t)=\delta(t) - \eta^2 {\cal D}^{(0)}_\eta(t)$
(with $\alpha$ set to 1).}
\label{fig:appendix-1}
\end{figure}

In this Appendix, to simplify the figures, we shall pictorially
denote the diffusons by single lines (with arrows), instead of double
lines as in Figs.\ \ref{fig:diagram-elements} and
\ref{fig:diagram-examples}. The order of the arrows at any
vertex remains important: it represents the
order of $b$-operators in the corresponding product
$\Tr (b{\bar b}b \dots {\bar b})$. In particular, arrows going
to/from any vertex alternate between incoming and outgoing directions.
Only vertices of even order are allowed. The diagram
from Fig.~\ref{fig:diagram-examples}c is again shown in Fig.~\ref{fig:appendix-1}a
in the new notation.  The vertices may contain differentiations
represented by a cross (Fig.~\ref{fig:appendix-1}b)
and $(\vp_i-\vp_j)^2$ terms (Fig.~\ref{fig:appendix-1}c).
When we transform a diffuson according to the equation of motion
(\ref{eq-motion}), there appear terms with the diffusons replaced
by the $\delta$-function. Such a replacement will be denoted
further as the double-crossed diffuson, Fig.~\ref{fig:appendix-1}d.

Before turning to canceling vertices, we calculate the numerical
prefactors at each diagram. For simplicity, we perform now the
calculation only for diagrams representing corrections to the
pro-diffuson (\ref{D-star-propagator}) defined as the
$\langle b{\bar b}\rangle$ propagator,
with one incoming and one outgoing diffusons.
Let $N_D$ denote the number of diffusons, $N_V$ the
number of vertices, and $2n_i$ be the vertex valencies.
Then $N_D=1+\sum n_i$ and the number of loops in the diagram
is $L=N_D-N_V-1$.

Every diffuson in the diagram brings in an additional factor of $(2/\pi)$.
In the rational parameterization (\ref{parameter-rational}),
collecting the coefficients of the vertices
$S_{\rm E}^{(\ge 4)}$ [Eq.~(\ref{E-vertex})]
and $S_{\rm kin}^{(\ge 4)}$ [Eqs.~(\ref{kin-vertex}) and (\ref{a-rational})]
(and taking into account the
$n_i$-fold symmetry of vertices (\ref{kin-vertex})), the total numerical
prefactor in front of the diagram for the pro-diffuson
${\cal D}^*=(\pi/2)\corr{b\bar b}$ [Eq.~(\ref{D-star-propagator})]
becomes $(-1)^{N_D-1}(2\pi)^{-L}$.
With this prefactor, each vertex coming from $S_{\rm E}^{(\ge 4)}$
enters with the coefficient 1, and each vertex originating
from $S_{\rm kin}^{(\ge 4)}$ enters with the coefficient
$\alpha a_{ij}$, where
\be
  a_{ij} = (-1)^{i-j+1} .
\label{a-ij}
\ee

Finally, by rescaling the time in the units of $\alpha^{-1/3}$, the
diagram acquires the overall factor $\alpha^{-L/3}$ ($\alpha^{1/3}$
for every vertex and $\alpha^{-(N_D-1)/3}$ from $N_D-1$ independent time
integrations). Now we may factor out the common coefficient for all the
diagrams of a given order $L$, and the remaining coefficients equal
plus or minus one. Note that within a given order $L$,
the relative sign alternates between diagrams with different number
of vertices.

\begin{figure}
\epsfxsize=0.5\hsize
\centerline{\epsfbox{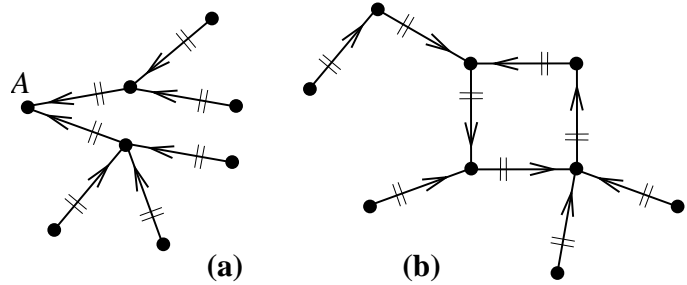}}
\medskip
\caption{
{\bf (a)} Tree-like graph of contracted diffusons.
{\bf (b)} Loop of contracted diffusons.}
\label{fig:appendix-2}
\end{figure}

The diagrammatic rules now may be formulated as follows.
\begin{enumerate}
\item
First we draw all the diagrams of a given order $L$ allowed by the
no-loop rule (see Section \ref{section-diagrams}).
The sign of the diagram is $(-1)^{N_D-1}$.

\item
In every vertex, we put the sum of all differentiations on outgoing
diffusons ($S_{\rm E}$-generated) and the full graph of
$a_{ij}(\varphi_i-\varphi_j)^2$ ($S_{\rm kin}$-generated),
see Fig.~\ref{fig:appendix-1}e.

\item
We expand every differentiation according to the diffuson equation
of motion (Fig.~\ref{fig:appendix-1}f).

\item
During this differentiation, some of the diffusons are replaced by
$\delta$ functions, which leads to merging or splitting some vertices
(discussed in details below).
Since every vertex has at most one differentiation on outgoing diffusons,
the connected graph of ``contracted'' diffusons contain no more than one
loop. We further distinguish two possibilities:
\begin{itemize}
\item[(i)]
the connected graph of ``contracted'' diffusons is a tree
(contains no loops), Fig.~\ref{fig:appendix-2}a. In this
case, all the vertices of the graph are contracted to a single
vertex (``root'' of the tree; point $A$ in Fig.~\ref{fig:appendix-2}a).
As the result of summing such diagrams, the coefficients
$a_{ij}$ at the ``root'' vertex get modified.

\item[(ii)]
the connected graph of ``contracted'' diffusons contains one loop,
Fig.~\ref{fig:appendix-2}b. Then upon contraction, we obtain two disconnected vertices without
any $(\varphi_i-\varphi_j)^2$ factors (vertices of this type would also be generated
by the Jacobian action as described in Appendix \ref{A:parameterizations},
if the Jacobian were not equal to one).
\end{itemize}
In the subsequent subsections we sum separately those two series of diagrams
and show that in the case (i) vertices of order higher than four cancel,
and in the case (ii) all such loop-contractions cancel each other.
\end{enumerate}

\begin{figure}
\epsfxsize=0.4\hsize
\centerline{\epsfbox{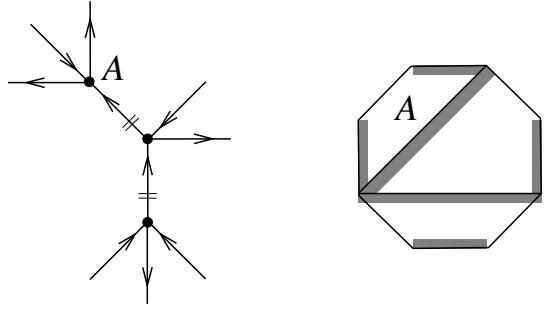}}
\medskip
\caption{
An example of a contraction tree and the corresponding polygon
(room-painting) diagram.}
\label{fig:appendix-3}
\end{figure}

\begin{figure}
\epsfxsize=0.6\hsize
\centerline{\epsfbox{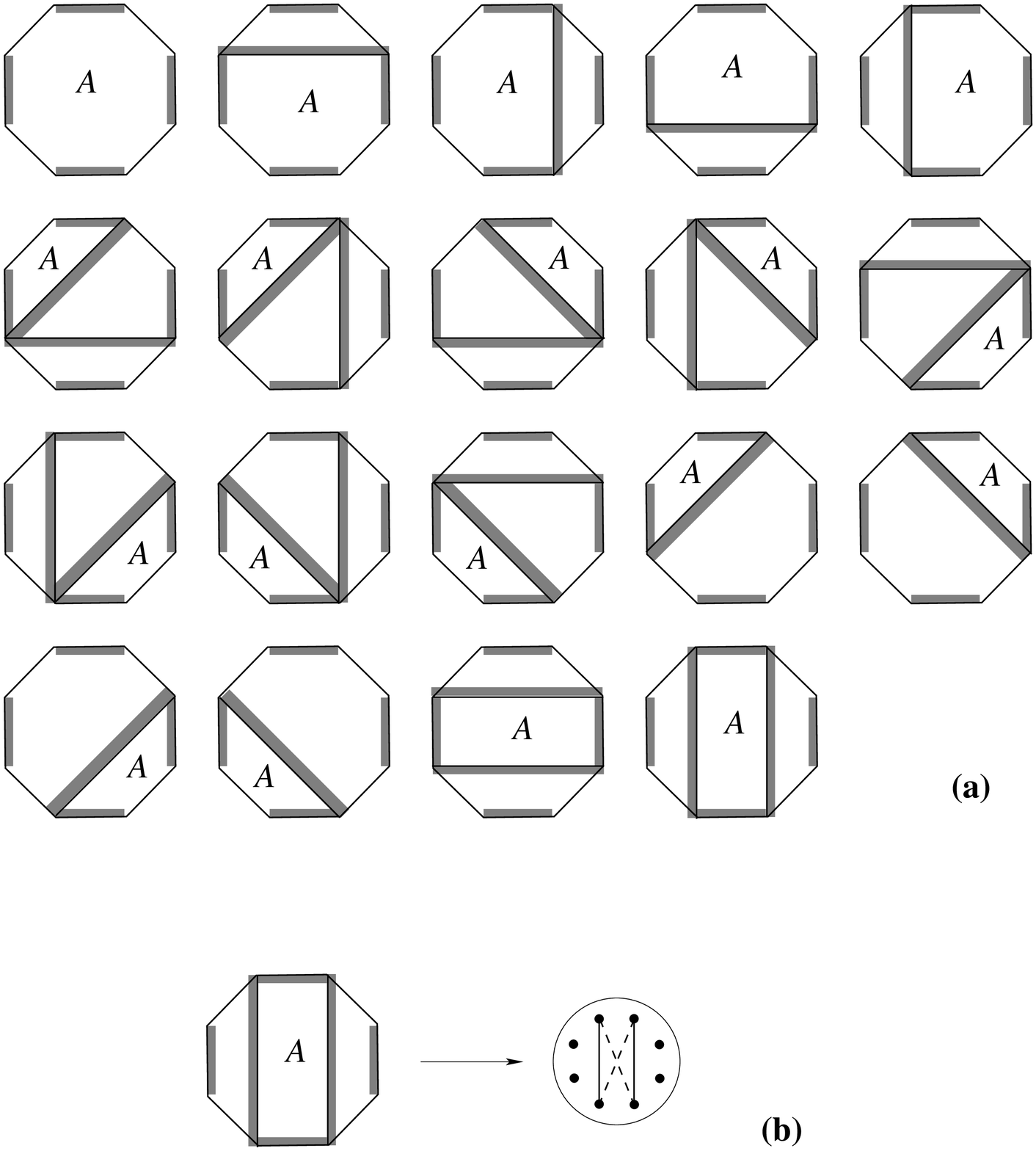}}
\medskip
\caption{
{\bf (a)} All admissible room partitions of the 8-wall room.
{\bf (b)} Every partition contributes the full graph of the ``distinguished''
room $A$ minus the external painted walls (and the whole graph is multiplied by
$(-1)$ to the number of internal walls).}
\label{fig:appendix-4a}
\end{figure}

\begin{figure}
\epsfxsize=0.6\hsize
\centerline{\epsfbox{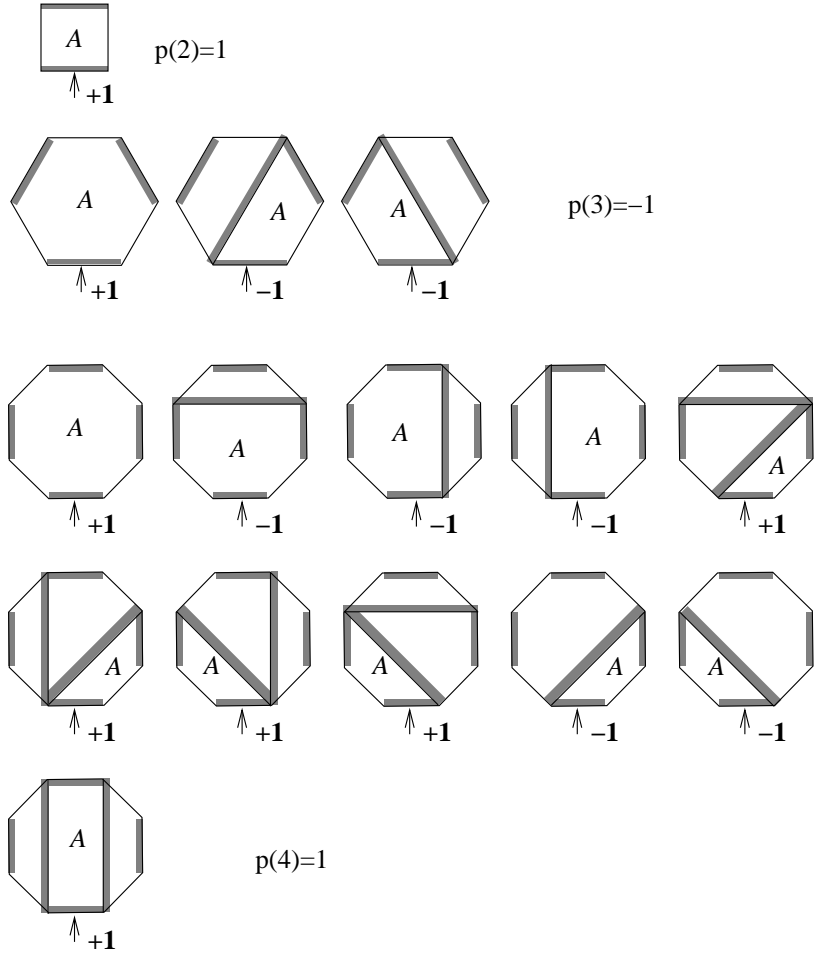}}
\medskip
\caption{
An example of a direct enumeration of $p(2)$, $p(3)$, $p(4)$.}
\label{fig:appendix-4b}
\end{figure}

\subsection{Cancellations in tree-like contractions}

Consider now the case when the connected graph of contracted diffusons
is a tree (Fig.~\ref{fig:appendix-2}a). This tree has one distinguished vertex
(denote it $A$) which does
not contain differentiation. All the $(\vp_i-\vp_j)^2$ terms of the resulting
``contracted'' vertex come from the vertex $A$. Now we represent the
contracted vertex as a $2n$-polygon, and the possible graph configurations
leading to this vertex as dividing this polygons by several diagonals into
smaller polygons (Fig.~\ref{fig:appendix-3}). Each such diagonal (a ``separator'')
should be thought of as a
``wall'' which represents a diffuson to be contracted.
We ``paint'' such a wall from the side of the $b$-operator (the side
of differentiation). The ``outside'' walls of the big $2n$-polygon are
also originally painted in the alternating order (Fig.~\ref{fig:appendix-3}). With this
representation, the different contractible graphs generating a given
vertex are in one-to-one correspondence with ``admissible'' partitions
of $2n$-polygons into smaller ``rooms''. A partition is called ``admissible''
if after painting all the walls in alternating order, no room has
more than one painted separator. Obviously, with this rule, exactly one
room of the partition has {\it no} painted separator. This room is
called the ``distinguished'' room of the partition. As an illustration,
in Fig.~\ref{fig:appendix-4a}a we show all the admissible partitions
of the 8-wall room.
From our discussion of the diagram rules and
prefactors, each partition contributes to the contracted vertex
the ``full graph of the distinguished
room'' minus ``painted walls of the distinguished room'' with the relative sign
of the parity number of rooms in the partition. Here the ``full graph''
means the full graph with the sign $\sum (-1)^{i-j+1}(\vp_i-\vp_j)^2$,
and the painted walls are subtracted when expanding the corresponding
differentiations (see fig.~\ref{fig:appendix-4a}b for an example).

Before we compute the total vertex as a sum of all room partitions,
we define a combinatoric quantity $p(n)$. Consider a $2n$-wall room with
external walls painted in the alternating order, and one of the painted walls
is ``special'' (say, it has a door). Let $p(n)$ denote the ``algebraic''
number of all admissible partitions
of the $2n$-room such that the distinguished room contains special
wall. By the ``algebraic'' number we mean that every partition
comes with plus or minus sign depending on the parity of the rooms in the
partition (and the trivial partition with only one room comes with plus sign).
For example, $p(2)=1$, $p(3)=-1$, $p(4)=1$ (Fig.~\ref{fig:appendix-4b}), etc.

\begin{figure}
\epsfxsize=0.7\hsize
\centerline{\epsfbox{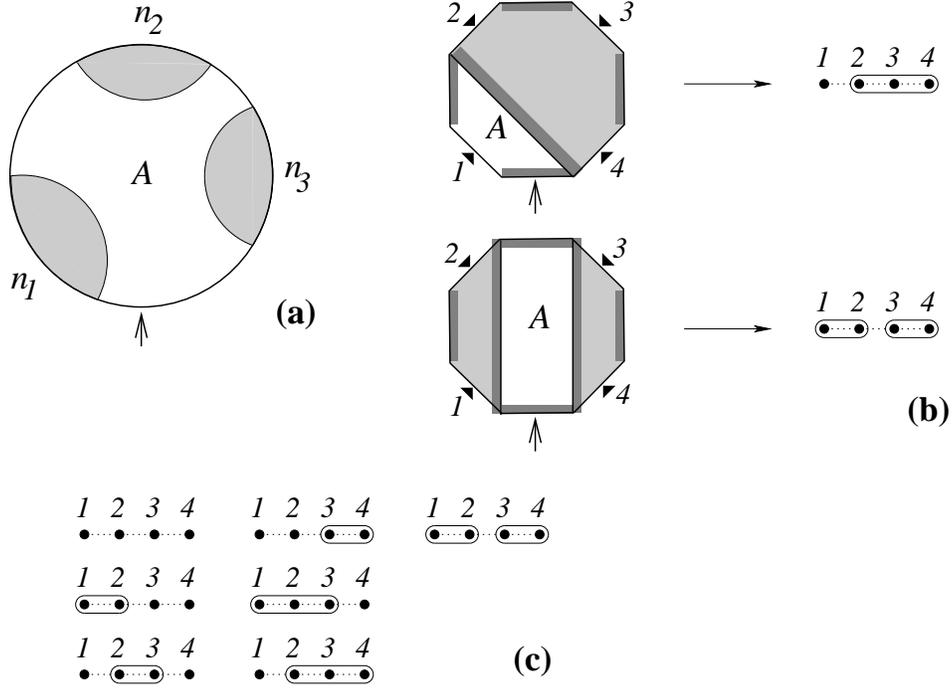}}
\medskip
\caption{
{\bf (a)} A schematic picture of external blocks adjacent
to the ``distinguished'' room A.
{\bf (b)} Combinatoric diagrams representing the external block
configurations.
{\bf (c)} All possible external block configurations in computing $p(4)$.
}
\label{fig:appendix-5}
\end{figure}

One can prove by induction that $p(n)=(-1)^n$. Indeed, all partitions
may be classified by ``external'' blocks
(shaded in fig.~\ref{fig:appendix-5}a) adjacent to
the distinguished room. Each of those external blocks contributes $-p(n_i)$
possibilities to subdivide it, where $n_i$ is the number of {\it unpainted}\/
external walls of the block. All possible configurations of external blocks
may be enumerated by dividing the total of $n$ unpainted external walls into
non-intersecting sequences of length $n_i$. Thus,
\be
p(n)=\sum \prod (-p(n_i)) ,
\label{pn-computation}
\ee
where the sum is taken over all possibilities to choose a set of
non-intersecting blocks of adjacent elements from a sequence of total $n$
elements (Fig.~\ref{fig:appendix-5}b). The only element excluded from the sum is the set where
the whole sequence belongs to a single block.
In Fig.~\ref{fig:appendix-5}c we illustrate the possible choices of such block sets
for $n=4$.
Now we perform the induction step assuming that for all $n_i<n$,
$p(n_i)=(-1)^{n_i}$. Then the same sum (\ref{pn-computation}) may be
rewritten as $p(n)=\sum (-1)^{n_{\rm adj}}$, where $n_{\rm adj}$ is the number
of adjacent pairs belonging to the same block. Finally, we note that
we may parameterize block sets by independently choosing two neighboring
elements either to belong or not to belong to the same block. This easily
gives us the total sum. If all the sets were allowed, the sum would be zero.
The only set which is not allowed is the one with all neighboring elements
(total $n-1$ pairs) belonging to the same block. This gives $p(n)=(-1)^n$,
which completes the proof.

\begin{figure}
\epsfxsize=0.5\hsize
\centerline{\epsfbox{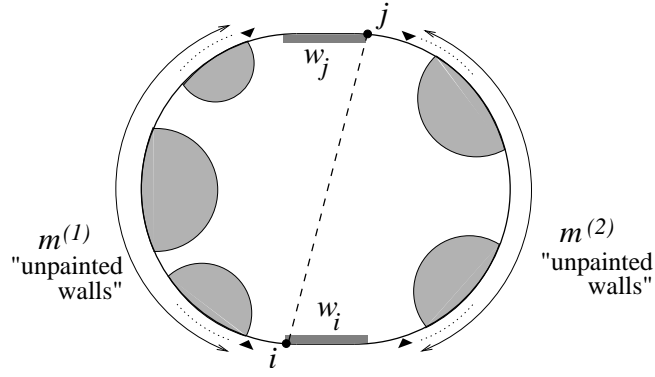}}
\medskip
\caption{The painted walls $w_i$ and $w_j$ divide the boundary into
two pieces of lengths $m^{(1)}$ and $m^{(2)}$.}
\label{fig:appendix-6}
\end{figure}

Now we do the final step of the calculation: we calculate the renormalized
coefficient $\tilde a_{ij}$
with which the combination $(\varphi_i-\varphi_j)^2$ enters the contracted vertex.
First, such terms with $i$ adjacent to $j$ through a {\it painted}\/ external
wall never enter (they are always subtracted by expanding differentiations).
So we may assume that $i$ and $j$ belong to different painted external walls
(let us call those walls $w_i$ and $w_j$ for future reference).

A room partition gives a contribution to $\tilde a_{ij}$ if and only if $w_i$
and $w_j$ both belong to the distinguished room.
Similarly to our discussion above, for each room partition,
we may consider external ``blocks'' complementary to the distinguished
room. Now the walls $w_i$ and $w_j$ divide the unpainted walls into two
sequences (define their lengths as $m^{(1)}$ and $m^{(2)}$),
and the contributions from these two sequences factorize:
\be
a_{ij}=(-1)^{i-j-1}
\sum_{(1)} \prod (-p(m^{(1)}_i)) \sum_{(2)} \prod (-p(m^{(2)}_i))
\ee
(see Fig.~\ref{fig:appendix-6}).
The only difference from the previous calculation is that now the
whole sequence $m^{(1)}$ (or $m^{(2)}$) is {\it allowed} to be placed in one
block. Therefore the sum gives zero unless {\it both} $m^{(1)}$ and $m^{(2)}$
equal one. But this is possible only if the total number of unpainted
external walls is $n=m^{(1)}+m^{(2)}=2$, i.e. for the four-valent vertex.
This completes the proof that all higher-order vertices cancel in the process
of tree-like contractions.

\begin{figure}
\epsfxsize=0.15\hsize
\centerline{\epsfbox{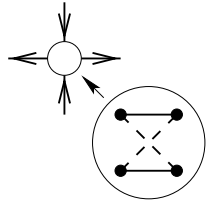}}
\medskip
\caption{
The resulting four-valent vertex.}
\label{fig:appendix-7}
\end{figure}

The only surviving four-leg vertex is shown in Fig.~\ref{fig:appendix-7},
with
\be
  \sum_{i<j} \tilde a_{ij} (\vp_i-\vp_j)^2
  = (\vp_2-\vp_3)^2 + (\vp_1-\vp_4)^2 - (\vp_1-\vp_3)^2 - (\vp_2-\vp_4)^2
  = 2 (\vp_1-\vp_2) (\vp_3-\vp_4) .
\ee
The diagrammatic series with such a vertex may be considered
as a perturbative expansion of the field theory
(\ref{S-star}), (\ref{b-action-rational}).

\subsection{Cancellation of loop-like contractions}

In this section we prove cancellation of vertices obtained in the
process of contracting a connected graph of diffusons containing
a closed loop (Fig.~\ref{fig:appendix-2}b). After such a contraction, two vertices appear
without any $(\phi_i-\phi_j)^2$ terms. This is the same type of vertices
as those appearing from the Jacobian. We consider the rational parameterization
which contains no Jacobian vertices, so the only vertices of this type
are those produced by the closed-loop diffuson contractions.

\begin{figure}
\epsfxsize=0.6\hsize
\centerline{\epsfbox{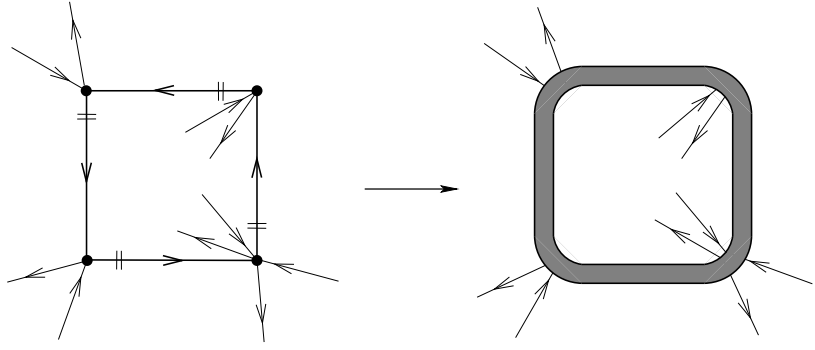}}
\medskip
\caption{
The loop contraction of diffusons. The inner and outer
boundaries of the shaded region in the right-hand side
are contracted into two separate vertices.}
\label{fig:appendix-8}
\end{figure}

Here we make two important remarks. First, there is no internal structure of those
vertices, and the only parameter of the vertex is its numerical prefactor.
Second, it is sufficient to prove cancellation of the ``loop part'' of the
diagram (Fig.~\ref{fig:appendix-8}) containing only the closed loop of contracted diffusons,
without any ``branches'' which do not depend on the internal structure
of the loop part.

\begin{figure}
\epsfxsize=0.7\hsize
\centerline{\epsfbox{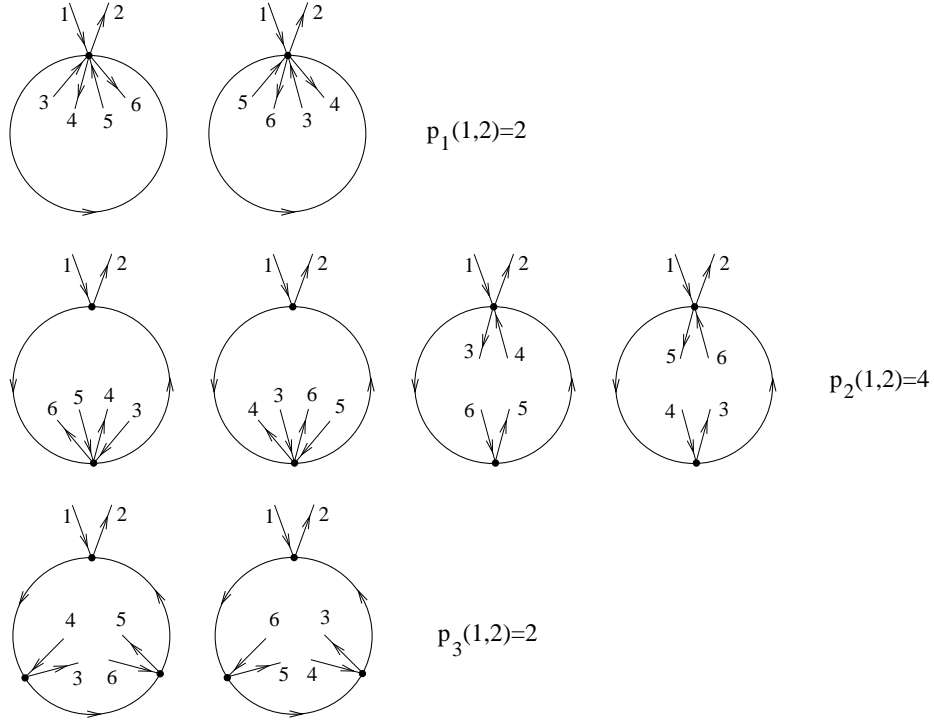}}
\medskip
\caption{
Direct computation of $p(1,2)=-p_1(1,2)+p_2(2,2)-p_3(1,2)=0$.}
\label{fig:appendix-9}
\end{figure}

So we consider loops of length $l\ge 1$ of contracted diffusons. If we go
along the loop, the legs on the left and on the right sides of our path
are collected into two new ``contracted'' vertices. We fix the valencies
of these vertices to be $2n$ and $2m$ and count the total combinatoric factor
$p(n,m)$ as the number of ways to obtain these vertices from loops of different
lengths. Taking into account the sign alternating with the total vertex
number, we obtain
\be
p(n,m)=\sum_{l=1}^{\infty} (-1)^l p_l(n,m) ,
\ee
where $p_l(n,m)$ is the positive integer number counting the number
of ways to create the two vertices of valencies $2n$ and $2m$ from the
loop of length $l$. Note that from the left-right hand rule both $n$ and
$m$ must be positive. Also, since the two-leg vertices are not allowed in
the original action, $p_l(n,m)=0$ for $l>n+m$. An example of calculating
$p(1,2)$ is shown in Fig.~\ref{fig:appendix-9}.

\begin{figure}
\epsfxsize=0.4\hsize
\centerline{\epsfbox{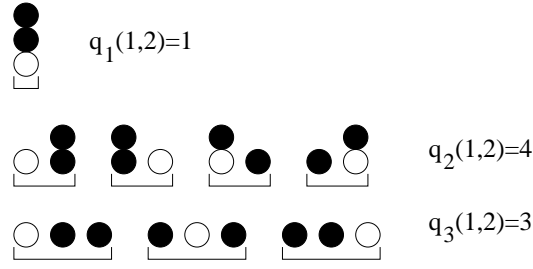}}
\medskip
\caption{
Example of $q_l(n,m)$ definition: direct computation of $q_l(1,2)$.}
\label{fig:appendix-10}
\end{figure}

The combinatoric problem of computing $p_l(n,m)$ may be related to
computing another combinatoric quantity
$q_l(n,m)$ defined as {\it the number of ways to distribute
$n$ identical black and $m$ identical white balls among $l$ different
boxes so that no box remains empty}:
\be
p_l(n,m)=n m \frac{1}{l} q_l(n,m)
\ee
(here factors $n$ and $m$ come from different ways to circularly
re-label external legs, and $1/l$ factors reflects equivalence of diagrams
obtained by the circular permutation of vertices). The example of
calculating $q_l(1,2)$ is shown in Fig.~\ref{fig:appendix-10}.

Finally, we use the generating-function method to compute $q_l(n,m)$.
Consider the series
\be
f_l(x,y)=(x+y+x^2+xy+y^2+\dots)^l ,
\ee
where the expression in the brackets contains all terms $x^{n_x} y^{n_y}$
except for $n_x=n_y=0$. Then the coefficient at $x^n y^m$ in the Taylor
expansion of $f(x,y)$ equals $q_l(n,m)$. Now a straightforward calculation
gives:
\be
f_l(x,y)=\left(\frac{1}{(1-x)(1-y)}-1\right)^l ,
\ee
and then
\be
\sum_{l=1}^{\infty} \frac{1}{l} (-1)^l f_l(x,y) = \ln (1-x) + \ln (1-y)
\label{generating_pl}
\ee
(note the factorization!). Now the coefficient $p(n,m)$ equals
$n m$ times the coefficient at $x^n y^m$ in the Taylor expansion of
the expression (\ref{generating_pl}). The latter however equals 0
unless one of the two numbers $n$ or $m$ is zero. This finishes
the proof: all vertices generated from loop contractions cancel.


\begin{thebibliography}{99}
\bibitem{GOE-comment}
For the Gaussian Orthogonal Ensemble,
the parameter $\alpha$ is defined as $\alpha=(\pi/2) C(0)v^2/\Delta^2$,
and $C(0)=\Delta^{-2}\overline{(\partial E_n/\partial\vp)^2}$.
\bibitem{convergence}
Convergence of the diagrams for $\Sigma_\eta(t)$ remains at the level
of conjecture. We explicitly see that the diagrams are convergent in
two- and four-loop calculations [integrals (\ref{K-sigma}) in the
four-loop case]. We do not have a proof of this conjecture at arbitrary
order.
\bibitem{UF-comment}
In notations of Ref.~\onlinecite{skvor} inherited from
Ref.~\onlinecite{KamenevAndreev99}, the lower-right block of $U_F$
was defined as $-\delta_{tt'}$. The two theories are trivially
equivalent by the rotation $U\mapsto\sigma_3U\sigma_3$
of the unitary matrices in Eq.~(\ref{orbit}).
\end{thebibliography}
\end{document}